\documentclass[english]{article}
\usepackage{geometry}
\usepackage{tikz}
\usepackage{amsmath,amssymb,graphicx,float}
\usepackage[unicode]{hyperref}
\usepackage{xcolor}
\usepackage{cite}
\usepackage{youngtab}

\allowdisplaybreaks[4]

\newcommand{\be}{\begin{equation}}
\newcommand{\ee}{\end{equation}}
\newcommand{\bea}{\begin{eqnarray}}
\newcommand{\eea}{\end{eqnarray}}
\newcommand{\ba}{\begin{array}}
\newcommand{\ea}{\end{array}}

\newcommand{\mb}{\mathbf}

\newcommand{\tr}{\mbox{Tr}}

\newcommand{\ud}{\underline}
\newtheorem{example}{Example}[section]

\begin{document}

\begin{titlepage}

\vspace*{5mm}%

\title{\textbf {A general fusion procedure for open $\mathfrak{gl}(N)$ spin chains: Application to the ABJM spin chain}}
	\author{Nan Bai$^{a}$\footnote{bainan@mailbox.gxnu.edu.cn}
}
	\date{}
{\let\newpage\relax\maketitle}
	\maketitle
	\underline{}
	\vspace{-10mm}
	
	\begin{center}
		{\it
            $^{a}$ Department of Physics, Guangxi Normal University, \\Guilin 541004, China
		}
		\vspace{10mm}
	\end{center}

\begin{abstract}
We formulate a general fusion procedure for open $\mathfrak{gl}(N)$ spin chains. We construct the fused boundary reflection matrices and the corresponding fused reflection equations. By using the intertwining relation between the fused reflection matrices and the fusion operator, we identify the invariant subspace of the fused reflection matrices carrying the irreducible representations of $\mathfrak{gl}(N)$.  We also construct the fused transfer matrix and evaluate it  explicitly in the total tensor product space and the invariant subspaces. Finally, we demonstrate that the ABJM spin chain model originates from such fusion procedure and derive three classes of boundary reflection matrices solutions on the anti-fundamental representation space of $\mathfrak{su}(4)$.
\end{abstract}
\end{titlepage}

\section{Introduction}

In many-body physics and quantum field theory, fusion refers to the formation of bound states from constituent particles. In Yang's pioneering work of quantum integrability\cite{Yang:1968ev}, the S-matrix for bound states has been constructed in terms of a series of ordered 2-2 scatterings  and proven to be both symmetric and unitary. Later, it was realized that factorizability is an intrinsic property in the (1+1)-dimensional integrable field theory\cite{Zamolodchikov:1978xm,Karowski:1978ps}, which indicates that the S-matrix for bound state can be fully determined by the S-matrices of fundamental particles. In the framework of algebraic Bethe ansatz, the S-matrices for bound-states correspond to the fused $R$-matrices which are solutions of the fused Yang-Baxter equation (YBE). Novel non-trivial fused $R$-matrices can be systematically constructed by the fusion procedure, which was first introduced in \cite{Kulish:1981gi,Kulish:1981bi} where the $R$-matrix for the spin-$\ell$ representation of $\mathfrak{su}(2)$ was obtained as the symmetrized tensor product of $2\ell$ fundamental spin-$\frac{1}{2}$ representations. Roughly speaking, the fusion procedure relies on the fact that, at some special values of the spectral parameter, the fundamental $R$-matrix reduces to the projection operator. When evaluated at these points, the fundamental YBE turns out to be an intertwining relation between the projector and the product of the remaining two $R$-matrices, and from this relation we obtain the fused $R$-matrix restricted to the projected subspace. Higher fused $R$-matrices can be constructed similarly from the fused YBE and the fusion operator. The fusion procedure has been extensively studied in two-dimensional solvable vertex and face models, including, for example: foundational work on solid-on-solid (SOS) model\cite{Date:1986ju,Date:1988}; a unified study for rational, trigonometric, and elliptic solutions in interaction-round-a-face (IRF) models\cite{Zhou:1989}; fusion hierarchies of $A–D–E$ face models\cite{Zhou:1994er}. For integrable models with boundaries, in addition to the fused $R$-matrices, one also needs to consider the fusion of boundary $K$-matrices, which are solutions to the reflection equation. The related fusion construction for integrable boundary models was developed in \cite{Mezincescu:1990fc,Mezincescu:1991ke,Zhou:1995zy}. For non-standard $R$-matrices appearing in AdS/CFT worldsheet scattering which do not reduce to projectors at special values of the spectral parameter, the corresponding fusion procedure is given in \cite{Beisert:2015msa} for fused $R$-matrices and \cite{Nepomechie:2015zwa} for fused $K$-matrices, respectively.

Beyond constructing new solutions, fusion techniques also provide essential functional relations for the exact solution of certain systems. Notably, they play a crucial role in the off-diagonal Bethe ansatz (ODBA) method developed in recent years \cite{Cao:2013nza,CYSW:2015}. ODBA method provides a systematic approach for handling integrable models with generic boundary conditions, particularly those lacking a reference state due to U(1) symmetry breaking. For higher-rank integrable models where the auxiliary space dimension exceeds 2 ($d>2$), additional functional equalities are required to express transfer matrices in closed form, thereby enabling the construction of nested T-Q relations that parameterize the eigenvalues of the fundamental transfer matrix. Crucially, the fusion method precisely provides these necessary functional relations:  For $\mathfrak{su}(n)$-invariant spin chains with generic integrable boundaries, the antisymmetric fusion operators have been used to construct the desired operator product identities\cite{Cao:2013cja}; for $\mathfrak{su}(2)$ higher spin models, the fusion hierarchy generated by symmetric fusion operators is employed to obatin the spectrum of the fundamental spin-($\frac{1}{2},s$) transfer matrix\cite{Cao:2014sta,Dong:2024ueo}. There are also applications in other types of Lie algebras (and their twisted versions) beyond $A_n$ including, for instance: the Izergin-Korepin model ($A_2^{(2)}$)\cite{Hao:2014fha}; the twisted $A_3^{(2)}$ model\cite{Li:2022cvg}; the $\mathfrak{so}(5)$ quantum spin chains ($B_2$)\cite{Li:2019cwb}; the $\mathfrak{sp}(4)$ quantum spin chain ($C_2$)\cite{Li:2018xrb}; the $C_n$ quantum spin chains\cite{Li:2020pen}; the twisted $C_2^{(1)}$ spin chains\cite{Li:2024fbh,Li:2024yqx}; the $\mathfrak{so}(6)$ quantum spin chains ($D_3^{(1)}$) \cite{Li:2019rzy} and its $q$-deformed extension \cite{Li:2023zdn}; the twisted $D_3^{(2)}$ model\cite{Li:2021qeq};  and also the quantum spin chains with exceptional Lie algebra $G_2$\cite{Li:2024uxd}.

In the fusion procedure, fused $R$-matrices are constructed on a tensor product space where the symmetric group acts naturally through permutation of tensor indices. When the fundamental $R$-matrix is taken to be the Yang solution\cite{Yang:1967bm}, the corresponding fusion operator can be shown to be the diagonal matrix element of an irreducible representation of the symmetric group.  Consequently, by the Schur-Weyl duality \cite{Schur-Weyl:1939}, the fused $R$-matrices restricted to irreducible representation subspaces of $\mathfrak{gl}(N)$ can be derived. The above relation between the fusion procedure and the representation theory of $\mathfrak{gl}(N)$ originates in the work of Jucys\cite{Jucys:1966} and has been applied in the quantum integrable system by Cherednik \cite{Cherednik:1989} and by Jimbo, Kuniba, Miwa and Okado\cite{Jimbo:1988gs}. The complete proofs can be found in \cite{Nazarov:1998,Molev:2008}. The fusion procedure can be generalized to irreducible representations of the Lie superalgebra $\mathfrak{gl}(N|M)$ using a super analogue of the classical Schur-Weyl duality \cite{Sergeev:1984,Berele:1987}, as well as to quantum groups via the Jimbo–Schur–Weyl duality between the Hecke algebra $H_n(q)$ and the quantum algebra $U_q(\mathfrak{gl}(N|M))$ \cite{Isaev:2008,Jimbo:1986,Moon:2001,Mitsuhashi:2005}. The related constructions are given in \cite{d'Andecy:2017}.

For closed spin chains, since the fused transfer matrix itself is a type of fused $R$-matrix, the study of the fused $R$-matrix and its invariant subspaces is sufficient for understanding the system, as detailed in \cite{Molev:2008,d'Andecy:2017}. However, for open spin chains, we must also fuse the boundary $K$-matrices and analyze their invariant subspaces. For $\mathfrak{su}(2)$ case, by means of the symmetric fusion operator, a concrete fusion procedure is presented in \cite{Zhou:1995zy} including the fused $K$-matrices and the fused transfer matrices along with their fusion hierarchy relations. A specific fusion procedure for $\mathfrak{sl}(N)$ chains, based on the existence of one-dimensional invariant subspace in the decomposition $\mathbf{1}\subset \mathbf{N}^{\otimes N}$, is given in \cite{Arnaudon:2004sd}. For general open $\mathfrak{gl}(N)$ spin chains, an explicit fusion scheme for arbitrary finite-dimensional irreducible representations of $\mathfrak{gl}(N)$ has not been established in the literature. The aim of this work is to bridge this gap: We will construct the fused $K^-$ and $K^+$ matrices (which yield the left and right boundary terms, respectively, of the open spin chain Hamiltonian), as well as the corresponding fused reflection and dual reflection equations. Crucially, we will identify the invariant subspaces of these fused $K$-matrices, which carry irreducible $\mathfrak{gl}(N)$ representations. Furthermore, we will construct the fused transfer matrices and evaluate them explicitly on these invariant subspaces. As an application, we will consider the ABJM spin chain model \cite{Minahan:2008hf,Bak:2008cp} whose quantum space alternates between the fundamental and anti-fundamental representations of $\mathfrak{su}(4)$.  A specific fusion model has been proposed in \cite{Bai:2024qdg}, where the fundamental and anti-fundamental representation spaces are fused into a single 16-dimensional quantum space. In the present work, we will study the fused $R$-matrix defined on four fundamental representation spaces. By applying antisymmetric fusion operator, we project the tensor product space of three of these spaces onto a four-dimensional anti-fundamental representation space. This allows us to re-derive the $R$-matrix given in the ABJM model, thereby demonstrating that the ABJM model constitutes a standard fusion model. We will also use the fusion procedure to explicitly construct three classes of $K^-$-matrix solutions on the anti-fundamental representation space.

The paper is organized as follows: In Section 2, we provide a comprehensive review of the fusion procedure for closed spin chains. By studying the intertwining relations between the fusion operator and the fused $R$-matrices, we identify the $\mathfrak{gl}(N)$ invariant subspaces of the fused $R$-matrices. In Section 3, we establish a general fusion procedure for open $\mathfrak{gl}(N)$ spin chains. We construct the fused $K^-$- and $K^+$-matrices and the corresponding fused reflection equation and dual fused reflection equation, respectively. The fused transfer matrices are then constructed and evaluated on both the total auxiliary space and the invariant subspaces. In Section 4, we demonstrate that the ABJM alternating spin chain model can be obtained by fusion on the $\mathfrak{su}(4)$ anti-fundamental representation space. We also derive some novel solutions for the $K^-$-matrices. In the last section, we make the conclusion and point out some future research directions. Finally, in the appendix, we provide another concrete example of the fusion method by obtaining the $R$-matrix of $\mathfrak{so}(6)$ model through the antisymmetric fusion of four $\mathfrak{su}(4)$ $R$-matrices.

\section{Fusion procedure for closed spin chains}

In this section, we review the fusion procedure for closed spin chains following basically the formulation in \cite{d'Andecy:2017}. We will investigate the fused $R$-matrices and the corresponding fused Yang-Baxter equation (YBE), and study the fusion operators as well as the invariant subspaces they generate. 

\begin{figure}[h]
 \begin{center}
   \includegraphics[width=1.0\linewidth]{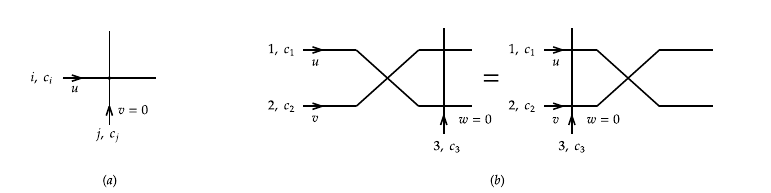}
 \end{center}
\caption{ (a) Fundamental $R$-matrix: $R_{ij}^{c_ic_j}(u-v)$;  (b) YBE with inhomogeneity parameters. \label{fig:rmybe}}
\end{figure}

To begin with, we recall the well-known fundamental YBE  defined on $V^{\otimes 3}$:
\begin{equation}\label{ybe}
R_{12}(u-v)R_{13}(u-w)R_{23}(v-w)=R_{23}(v-w)R_{13}(u-w)R_{12}(u-v)\quad u,v,w \in\mathbb{C},
\end{equation}
with the fundamental $R$-matrix $R_{12}(u,v)=R_{12}(u-v)\in {\rm{End}}(V\otimes V)$. A crucial observation here: in addition to the spectral variables $u,v,w$, we may also assign a constant inhomogeneity parameter $c_i$ to each space $V_i$ , and the corresponding $R$-matrix is denoted as:
\begin{equation}\label{eqno1}
R^{c_ic_j}_{ij}(u):=R_{ij}(u+c_i-c_j),
\end{equation}
which is the solution of the following type YBE:
\begin{equation}\label{eqno2}
R^{c_1c_2}_{12}(u-v)R_{13}^{c_1c_3}(u)R_{23}^{c_2c_3}(v)=R_{23}^{c_2c_3}(v)R_{13}^{c_1c_3}(u)R^{c_1c_2}_{12}(u-v).
\end{equation}
The graphical representations for $R$-matrix (\ref{eqno1}) and YBE (\ref{eqno2}) with inhomogeneities are plotted in Fig.~\ref{fig:rmybe} (a) and (b) , respectively. Such geometrical constructions for algebraic relations will be extensively employed throughout the paper.

Remark: The arrows in the above figure indicate the moving directions of the particles. When two particles meet at a crossing point, we assign a corresponding $R$-matrix, which can be regarded as a two-particle scattering matrix. Consequently, the arrows uniquely determine the temporal sequence of the scattering events, and hence the order in which the corresponding $R$-matrices are multiplied.

We will consider exclusively the Yang solution \cite{Yang:1967bm} of (\ref{ybe}) :
\begin{equation}\label{yangsl}
R_{ij}(u)=\mathbb{I}-\frac{\mathbb{P}_{ij}}{u}\in{\rm{End}}(V_i\otimes V_j),
\end{equation}
which is the representation of the element $(1)-\frac{(i,j)}{u}$ of the group algebra $\mathbb{C}[S_n]$, where $(1)$ denotes the identity and $(i,j)$ the permutation element in the symmetric group $S_n$. The reason for using Yang solution stems from the property that, as will be demonstrated below, the fusion operators constructed from Yang $R$-matrices yield projection operators which give fused $R$-matrices on irreducible $\mathfrak{gl}(N)$-subspaces.
\subsection{Fused $R$-matrix and fused Yang-Baxter equation}
Now we investigate the fused $R$-matrix, which physically corresponds to the S-matrix for bound states of composite particles. 
Concretely, let us consider the scattering between an $n$-particle bound state and an $n'$-particle bound state, whose fused $R$-matrix is denoted as
\begin{equation}\label{frm}
R_{(1\cdots n),(\ud{1}\cdots \ud{n}')}^{\mathbf{c}_{(1\cdots n)},\mathbf{c}_{(\ud{1}\cdots\ud{n}')}}(u-v),
\end{equation}
where
\begin{itemize}
  \item For $n$-particle bound state $(1\cdots n)$: All constituent particles share the same spectral parameter $u$ but may possess distinct inhomogeneity parameters $\mb{c}_{(1\cdots n)} := \{c_1,\dots,c_n\}$; 
  \item For $n'$-particle bound state $(\ud{1}\cdots\ud{n}')$: All constituent particles have the same spectral parameter $v$ and possibly distinct inhomogeneities $\mb{c}_{(\ud{1}\cdots\ud{n}')}=\{c_{\ud{1}}\cdots c_{\ud{n}'}\}$.
\end{itemize}

\begin{figure}[h]
 \begin{center}
   \includegraphics[width=0.7\linewidth]{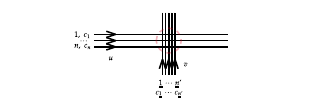}
 \end{center}
\caption{ Graphical representation for the fused $R$-matrix$: R_{(1\cdots n),(\ud{1}\cdots \ud{n}')}^{\mathbf{c}_{(1\cdots n)},\mathbf{c}_{(\ud{1}\cdots\ud{n}')}}(u-v)$.\label{fig:frm}}
\end{figure} 

The fused $R$-matrix (\ref{frm}) is graphically shown in Fig.~(\ref{fig:frm}). The dashed circle in the diagram indicates that the enclosed area, as a whole, denotes a related quantity (in this case, the fused $R$-matrix), and so hereinafter.

The explicit expression of the fused $R$-matrix can be directly obtained from Fig.~(\ref{fig:frm}): by assigning each crossing point with the corresponding fundamental $R$-matrix and multiplying these fundamental $R$-matrices in an appropriate order. The concrete result is as follows:
\begin{equation}\label{frmxp}
\begin{split}
R_{(1\cdots n),(\ud{1}\cdots \ud{n}')}^{\mathbf{c}_{(1\cdots n)},\mathbf{c}_{(\ud{1}\cdots\ud{n}')}}(u)
=&\left[R_{n\ud{1}}^{c_n c_{\ud{1}}}(u)R_{n\ud{2}}^{c_n c_{\ud{2}}}(u)\cdots R_{n\ud{n}'}^{c_n c_{\ud{n}'}}(u)\right]
\cdots
\left[R_{1\ud{1}}^{c_1 c_{\ud{1}}}(u)R_{1\ud{2}}^{c_1 c_{\ud{2}}}(u)\cdots R_{1\ud{n}'}^{c_1 c_{\ud{n}'}}(u)\right]\\
=&\overset{\leftarrow}{\prod_{i=1\cdots n}}\left[R_{i\ud{1}}^{c_i c_{\ud{1}}}(u)R_{i\ud{2}}^{c_i c_{\ud{2}}}(u)\cdots R_{i\ud{n}'}^{c_i c_{\ud{n}'}}(u)\right]\\
=&\overset{\rightarrow}{\prod_{i=1\cdots n'}}\left[R_{n\ud{i}}^{c_nc_{\ud{i}}}(u)R_{n-1,\ud{i}}^{c_{n-1},c_{\ud{i}}}(u)\cdots R_{1\ud{i}}^{c_1c_{\ud{i}}}(u) \right],
\end{split}
\end{equation}
where $\overset{\rightarrow}{\prod}_i$ and $\overset{\leftarrow}{\prod}_i$ denote ordered products taken with increasing (lexicographical) and decreasing (reverse lexicographical) indices, respectively \footnote{We note that an expression similar to (\ref{frmxp}) has appeared in the original paper \cite{Yang:1968ev} ( Eq.~(8)) as the S-matrix for bound states.}. 
 
\begin{figure}[h]
 \begin{center}
   \includegraphics[width=1.0\linewidth]{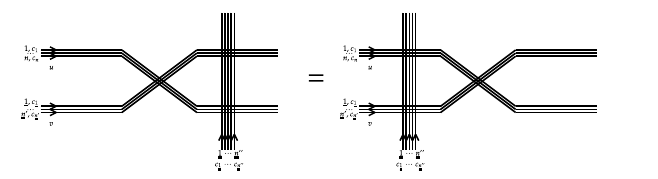}
 \end{center}
\caption{ Graphical representation of the fused YBE. \label{fig:fYBE}}
\end{figure}

 We proceed to study the fused YBE which reflects the factorized three-body scattering process of composite particles. Its graphical representation, analogous to that of the fundamental YBE, is depicted in Fig.~\ref{fig:fYBE}, with the concrete expression given by:
 \begin{equation}\label{fYBE}
\begin{split}
&R_{(1\cdots n),(\ud{1}\cdots \ud{n}')}^{\mathbf{c}_{(1\cdots n)},\mathbf{c}_{(\ud{1}\cdots\ud{n}')}}(u-v)
R_{(1\cdots n),(\ud{\ud{1}}\cdots\ud{\ud{n}}'')}^{\mathbf{c}_{(1\cdots n)},\mathbf{c}_{(\ud{\ud{1}}\cdots \ud{\ud{n}}'')}}(u)
R_{(\ud{1}\cdots \ud{n}'),(\ud{\ud{1}}\cdots\ud{\ud{n}}'')}^{\mathbf{c}_{(\ud{1}\cdots \ud{n}')},\mathbf{c}_{(\ud{\ud{1}}\cdots \ud{\ud{n}}'')}}(v)\\
=&R_{(\ud{1}\cdots \ud{n}'),(\ud{\ud{1}}\cdots\ud{\ud{n}}'')}^{\mathbf{c}_{(\ud{1}\cdots \ud{n}')},\mathbf{c}_{(\ud{\ud{1}}\cdots \ud{\ud{n}}'')}}(v)
R_{(1\cdots n),(\ud{\ud{1}}\cdots\ud{\ud{n}}'')}^{\mathbf{c}_{(1\cdots n)},\mathbf{c}_{(\ud{\ud{1}}\cdots \ud{\ud{n}}'')}}(u)
R_{(1\cdots n),(\ud{1}\cdots \ud{n}')}^{\mathbf{c}_{(1\cdots n)},\mathbf{c}_{(\ud{1}\cdots\ud{n}')}}(u-v).
\end{split}
\end{equation}
Based on the fundamental YBE (\ref{eqno2}) and the definition of the fused $R$-matrix (\ref{frmxp}), the fused YBE (\ref{fYBE}) can be rigorously proved  by mathematical induction (see \cite{d'Andecy:2017} for details).

\subsubsection{Some properties of the fused $R$-matrices}
We present some properties of the fused $R$-matrix, which will be frequently used in subsequent derivations.
\begin{itemize}
\item [(1)] P-symmetry: First note that for the fundamental $R$-matrix $R^{c_ic_j}_{ij}(u)$, we have:
\begin{equation}
R_{ij}^{c_ic_j}(u)=\mathbb{I}-\frac{\mathbb{P}_{ij}}{u+c_i-c_j}=\mathbb{I}-\frac{\mathbb{P}_{ji}}{u+(-c_j)-(-c_i)}=R_{ji}^{-c_j,-c_i}(u),
\end{equation} 
which gives
\begin{equation}
\begin{split}
R_{i,(\ud{1}\cdots\ud{n}')}^{c_i,\mb{c}_{(\ud{1}\cdots\ud{n}')}}(u)=R_{\ud{1}i}^{-c_{\ud{1}},-c_i}(u)\cdots R_{\ud{n}',i}^{-c_{\ud{n}'},-c_i}(u)
=R_{(\ud{n}'\cdots\ud{1}),i}^{-\mb{c}_{(\ud{n}'\cdots\ud{1})},-c_i}(u).
\end{split}
\end{equation}
Thus,
\begin{equation}\label{psym}
\begin{split}
R_{(1\cdots n),(\ud{1}\cdots\ud{n}')}^{\mb{c}_{(1\cdots n)},\mb{c}_{(\ud{1}\cdots\ud{n}')}}(u)
=&R_{n,(\ud{1}\cdots\ud{n}')}^{c_n,\mb{c}_{\ud{1}\cdots\ud{n}'}}(u)\cdots R_{1,(\ud{1}\cdots\ud{n}')}^{c_1,\mb{c}_{\ud{1}\cdots\ud{n}'}}(u)\\
=&R_{(\ud{n}'\cdots\ud{1}),n}^{-\mb{c}_{(\ud{n}'\cdots\ud{1})},-c_n}(u)\cdots R_{(\ud{n}'\cdots\ud{1}),1}^{-\mb{c}_{(\ud{n}'\cdots\ud{1})},-c_1}(u)
=R_{(\ud{n}'\cdots\ud{1}),(n\cdots 1)}^{-\mb{c}_{(\ud{n}'\cdots \ud{1})},-\mb{c}_{(n\cdots 1)}}(u),
\end{split}
\end{equation}
which shows the P-symmetry possessed by the fused $R$-matrix. 
\item [(2)] T-symmetry: We will use $t_i$ to denote the transpose in the space $V_i$ and $t_{(1\cdots n)}=t_1t_2\cdots t_n$ the transpose in the tensor product space: $V_{(1\cdots n)}:=V_1\otimes V_2\otimes\cdots\otimes V_n$. Since $R_{ij}^{c_ic_j}(u)^{t_i}=R_{ij}^{c_ic_j}(u)^{t_j}$, we have:
    \begin{equation}\label{tsym1}
    \begin{split}
    &R_{(1\cdots n),\ud{i}}^{\mb{c}_{(1\cdots n)},c_{\ud{i}}}(u)^{t_{(1\cdots n)}}
    =\left[R_{n\ud{i}}^{c_nc_{\ud{i}}}(u)\cdots R_{1\ud{i}}^{c_1c_{\ud{i}}}(u)\right]^{t_{(1\cdots n)}}
    =\left[R_{n\ud{i}}^{c_nc_{\ud{i}}}(u)^{t_n}\cdots R_{1\ud{i}}^{c_1c_{\ud{i}}}(u)^{t_1}\right]\\
    =&\left[R_{n\ud{i}}^{c_nc_{\ud{i}}}(u)^{t_{\ud{i}}}\cdots R_{1\ud{i}}^{c_1c_{\ud{i}}}(u)^{t_{\ud{i}}}\right]
    =\left[R_{1\ud{i}}^{c_1c_{\ud{i}}}(u)\cdots R_{n\ud{i}}^{c_nc_{\ud{i}}}(u)\right]^{t_{\ud{i}}}
    =R_{(n\cdots 1),\ud{i}}^{\mb{c}_{(n\cdots 1)},c_{\ud{i}}}(u)^{t_{\ud{i}}} ,
    \end{split}
    \end{equation}
    where we have used the fact: $(A\otimes B)^{t_At_B}=A^{t_A}\otimes B^{t_B}$. Then we have:
    \begin{equation}\label{tsym2}
    \begin{split}
    R_{(1\cdots n),(\ud{1}\cdots\ud{n}')}^{\mb{c}_{(1\cdots n)},\mb{c}_{(\ud{1}\cdots\ud{n}')}}(u)^{t_{(1\cdots n)}}=&
    \left[R_{(1\cdots n),\ud{1}}^{\mb{c}_{(1\cdots n)},c_{\ud{1}}}(u)\cdots R_{(1\cdots n),\ud{n}'}^{\mb{c}_{(1\cdots n)},c_{\ud{n}'}}(u)\right]^{t_{(1\cdots n)}}\\
    =&R_{(1\cdots n),\ud{n}'}^{\mb{c}_{(1\cdots n)},c_{\ud{n}'}}(u)^{t_{(1\cdots n)}}\cdots R_{(1\cdots n),\ud{1}}^{\mb{c}_{(1\cdots n)},c_{\ud{1}}}(u)^{t_{(1\cdots n)}}\\
    =&R_{(n\cdots 1),\ud{n}'}^{\mb{c}_{(n\cdots 1)},c_{\ud{i}}}(u)^{t_{\ud{n}'}}\cdots R_{(n\cdots 1),\ud{1}}^{\mb{c}_{(n\cdots 1)},c_{\ud{1}}}(u)^{t_{\ud{1}}}=R_{(n\cdots 1),(\ud{n}'\cdots\ud{1})}^{\mb{c}_{(n\cdots 1)},\mb{c}_{(\ud{n}'\cdots\ud{1})}}(u)^{t_{(\ud{n}'\cdots \ud{1})}},
    \end{split}
    \end{equation}
    which is the T-symmetry satisfied by the fused $R$-matrix. This relation can also be written as:
    \begin{equation}\label{tsym3}
    R_{(1\cdots n),(\ud{1}\cdots\ud{n}')}^{\mb{c}_{(1\cdots n)},\mb{c}_{(\ud{1}\cdots\ud{n}')}}(u)^{t_{(1\cdots n)},t_{(\ud{1}\cdots\ud{n}')}}
    =R_{(n\cdots 1),(\ud{n}'\cdots\ud{1})}^{\mb{c}_{(n\cdots 1)},\mb{c}_{(\ud{n}'\cdots\ud{1})}}(u).
    \end{equation}
    
\item [(3)] Unitary condition: The fundamental $R$-matrix satisfies the following unitary condition:
\begin{equation}
R_{ij}^{c_ic_j}(u)R_{ji}^{c_jc_i}(-u)=\left(\mathbb{I}-\frac{\mathbb{P}_{ij}}{u+c_i-c_j}\right)\left(\mathbb{I}-\frac{\mathbb{P}_{ji}}{-u+c_j-c_i}\right)
=\left[1-\frac{1}{(u+c_i-c_j)^2}\right]\mathbb{I},
\end{equation}
which implies
\begin{equation}
\begin{split}
R_{i,(\ud{1}\cdots\ud{n}')}^{c_i,\mb{c}_{(\ud{1}\cdots\ud{n}')}}(u)R_{(\ud{1}\cdots\ud{n}'),i}^{\mb{c}_{(\ud{1}\cdots\ud{n}')},c_i}(-u)=&
\left[R_{i\ud{1}}^{c_ic_{\ud{1}}}(u)\cdots R_{i\ud{n}'}^{c_ic_{\ud{n}'}}(u)\right]\left[R_{\ud{n}'i}^{c_{\ud{n}'}c_i}(-u)\cdots R_{\ud{1} i}^{c_{\ud{1}}c_i}(-u)\right]\\
=&\prod_{j=1}^{n'}\left[1-\frac{1}{(u+c_i-c_{\ud{j}})^2}\right]\mathbb{I}.
\end{split}
\end{equation}
Consequently, we derive the unitary condition for the fused $R$-matrix:
\begin{equation}\label{fuc}
\begin{split}
&R_{(1\cdots n),(\ud{1}\cdots\ud{n}')}^{\mb{c}_{(1\cdots n)},\mb{c}_{(\ud{1}\cdots\ud{n}')}}(u)
R_{(\ud{1}\cdots\ud{n}'),(1\cdots n)}^{\mb{c}_{(\ud{1}\cdots\ud{n}')},\mb{c}_{(1\cdots n)}}(-u)\\
=&\left[R_{n,(\ud{1}\cdots\ud{n}')}^{c_n,\mb{c}_{(\ud{1}\cdots\ud{n}')}}(u)\cdots R_{1,(\ud{1}\cdots\ud{n}')}^{c_1,\mb{c}_{(\ud{1}\cdots\ud{n}')}}(u)\right]
\left[R_{(\ud{1}\cdots\ud{n}'),1}^{\mb{c}_{(\ud{1}\cdots\ud{n}')},c_1}(-u)\cdots 
R_{(\ud{1}\cdots\ud{n}'),n}^{\mb{c}_{(\ud{1}\cdots\ud{n}')},c_n}(-u)\right]\\
=&\prod_{i=1}^{n}\prod_{j=1}^{n'}\left[1-\frac{1}{(u+c_i-c_{\ud{j}})^2}\right]\mathbb{I}.
\end{split}
\end{equation}
For later convenience, we introduce the symbol $\rho_{(1\cdots n),(\ud{1}\cdots\ud{n}')}^{\mb{c}_{(1\cdots n)},\mb{c}_{(\ud{1}\cdots\ud{n}')}}(u)$ to denote the scalar function in the above result (\ref{fuc}): $\rho_{(1\cdots n),(\ud{1}\cdots\ud{n}')}^{\mb{c}_{(1\cdots n)},\mb{c}_{(\ud{1}\cdots\ud{n}')}}(u)=\prod_{i=1}^{n}\prod_{j=1}^{n'}\left[1-\frac{1}{(u+c_i-c_{\ud{j}})^2}\right]$.

\item [(4)] Crossing unitary condition: For the fundamental $R$-matrix acting on $V\otimes V$ with $N={\rm{dim}}V$, the crossing unitary condition is given by:
\begin{equation}
R_{ij}^{c_ic_j}(u)^{t_i}R_{ji}^{c_jc_i}(-u+N)^{t_i}=\mathbb{I}.
\end{equation}
Utilizing relations (\ref{tsym1}) and (\ref{tsym2}), we derive the partial transposes for the fused $R$-matrix:
\begin{equation}
\begin{split}
&R_{(1\cdots n),(\ud{1}\cdots\ud{n}')}^{\mb{c}_{(1\cdots n)},\mb{c}_{(\ud{1}\cdots\ud{n}')}}(u)^{t_{(1\cdots n)}}
=\left[R_{n\ud{n}'}^{c_nc_{\ud{n}'}}(u)^{t_n}\cdots R_{1\ud{n}'}^{c_1c_{\ud{n}'}}(u)^{t_1}\right]\cdots
\left[R_{n\ud{1}}^{c_nc_{\ud{1}}}(u)^{t_n}\cdots R_{1\ud{1}}^{c_1c_{\ud{1}}}(u)^{t_1}\right],\\
&R_{(\ud{1}\cdots\ud{n}'),(1\cdots n)}^{\mb{c}_{(\ud{1}\cdots\ud{n}')},\mb{c}_{(1\cdots n)}}(u)^{t_{(1\cdots n)}}
=\left[R_{\ud{1}1}^{c_{\ud{1}}c_1}(u)^{t_1}\cdots R_{\ud{1}n}^{c_{\ud{1}}c_n}(u)^{t_n}\right]
\cdots
\left[R_{\ud{n}'1}^{c_{\ud{n}'}c_1}(u)^{t_1}\cdots R_{\ud{n}'n}^{c_{\ud{n}'}c_n}(u)^{t_n}\right].
\end{split}
\end{equation}
Consequently, the crossing unitary condition for the fused $R$-matrix takes the form:
\begin{equation}\label{fcuc}
R_{(1\cdots n),(\ud{1}\cdots\ud{n}')}^{\mb{c}_{(1\cdots n)},\mb{c}_{(\ud{1}\cdots\ud{n}')}}(u)^{t_{(1\cdots n)}}
R_{(\ud{1}\cdots\ud{n}'),(1\cdots n)}^{\mb{c}_{(\ud{1}\cdots\ud{n}')},\mb{c}_{(1\cdots n)}}(-u+N)^{t_{(1\cdots n)}}=\mathbb{I}.
\end{equation}

\end{itemize}
\subsection{Invariant subspaces of the fused $R$-matrices}
The tensor product space $V^{\otimes n}\otimes V^{\otimes n'}$ for the fused $R$-matrix $R_{(1\cdots n),(\ud{1}\cdots\ud{n}')}^{\mb{c}_{(1\cdots n)},\mb{c}_{(\ud{1}\cdots\ud{n}')}}(u)$ is generally too large for practical calculations. We therefore seek to identify invariant subspaces, particularly those carrying irreducible $\mathfrak{gl}(N)$ representations, to obtain the reduced form of the fused $R$-matrix. For this purpose, we will introduce the fusion operators in this section. When configured with appropriate inhomogeneity parameters, these operators will correspond to the primitive idempotents in the $\mathbb{C}[S_n]$-module, and serve as the required projection operators onto $\mathfrak{gl}(N)$ irreducible subspaces. We will also establish the intertwining relations between the fused $R$-matrices and the fusion operators, and examine  some other properties of the latter.

\subsubsection{General intertwining relations}
\begin{figure}[h]
 \begin{center}
   \includegraphics[width=0.65\linewidth]{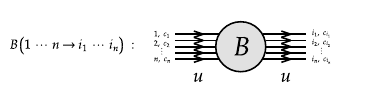}
 \end{center}
\caption{Braiding operator $B(1\cdots n\rightarrow i_1\cdots i_n)$. \label{fig:io}}
\end{figure}

We first introduce the \textit{braiding operator} $B(1\cdots n\rightarrow i_1\cdots i_n)$ : it describes the specific planar scattering process of 
$n$ particles from initial configuration $(1\cdots n)$ to final configuration $(i_1\cdots i_n)$. All particles share common spectral parameter $u$ with distinct inhomogeneities $\{c_1\cdots c_n\}$, as schematized in Fig.~\ref{fig:io}.

\begin{figure}[h]
 \begin{center}
   \includegraphics[width=0.65\linewidth]{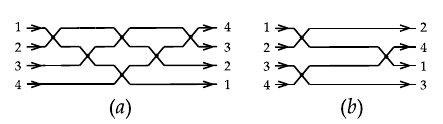}
 \end{center}
\caption{Two specific 4-particle braiding operators: $(a): B(1234\rightarrow 4321)$ $(b): B(1234\rightarrow 2413)$.  \label{fig:2iw}}
\end{figure}

The explicit form of the braiding operator depends on the specific scattering process. For instance, Fig.~\ref{fig:2iw} illustrates two distinct 4-particle scattering processes: $(1234\rightarrow 4321)$ and $(1234\rightarrow 2413)$, with their concrete expressions given by:
\begin{eqnarray}
&&B(1234\rightarrow 4321)=R_{12}^{c_1c_2}(0)R_{13}^{c_1c_3}(0)R_{14}^{c_1c_4}(0)R_{23}^{c_2c_3}(0)R_{24}^{c_2c_4}(0)R_{34}^{c_3c_4}(0),\\
&&B(1234\rightarrow 2413)=R_{12}^{c_1c_2}(0)R_{34}^{c_3c_4}(0)R_{14}^{c_1c_4}(0).
\end{eqnarray}

\begin{figure}[h]
 \begin{center}
   \includegraphics[width=0.7\linewidth]{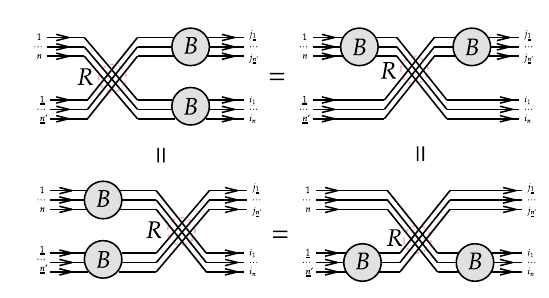}
 \end{center}
\caption{Graphical representation of ``$RBB$" relations.  \label{fig:RBB}}
\end{figure}

The fused $R$-matrices and the braiding operators satisfy several equivalent intertwining relations (to be called ``$RBB$" relations), which are graphically represented in Fig.~\ref{fig:RBB}, with the explicit expressions:
\begin{equation}\label{RBB}
\begin{split}
&R_{(1\cdots n),(\ud{1}\cdots\ud{n}')}^{\mb{c}_{(1\cdots n)},\mb{c}_{(\ud{1}\cdots\ud{n}'}}(u-v)B(1\cdots n\rightarrow i_1\cdots i_n)B(\ud{1}\cdots\ud{n}'\rightarrow j_{\ud{1}}\cdots j_{\ud{n}'})\\
=&B(1\cdots n\rightarrow i_1\cdots i_n)R_{(i_1\cdots i_n),(\ud{1}\cdots\ud{n}')}^{\mb{c}_{(i_1\cdots i_n)},\mb{c}_{(\ud{1}\cdots\ud{n}')}}(u-v)B(\ud{1}\cdots\ud{n}'\rightarrow j_{\ud{1}}\cdots j_{\ud{n}'})\\
=&B(\ud{1}\cdots\ud{n}'\rightarrow j_{\ud{1}}\cdots j_{\ud{n}'})
R_{(1\cdots n),(j_{\ud{1}}\cdots j_{\ud{n}'})}^{\mb{c}_{(1\cdots n)},\mb{c}_{(j_{\ud{1}}\cdots j_{\ud{n}'})}}(u-v)B(1\cdots n\rightarrow i_1\cdots i_n)\\
=&B(1\cdots n\rightarrow i_1\cdots i_n)B(\ud{1}\cdots\ud{n}'\rightarrow j_{\ud{1}}\cdots j_{\ud{n}'})
R_{(i_1\cdots i_n),(j_{\ud{1}}\cdots j_{\ud{n}'})}^{\mb{c}_{(i_1\cdots i_n)},\mb{c}_{j_{\ud{1}}\cdots j_{\ud{n}'}}}(u-v).
\end{split}
\end{equation}

The above $RBB$ relations can be understood diagrammatically: As shown in Fig.~\ref{fig:RBB}, by sequentially moving the single-particle lines across the braiding operator by means of the fundamental YBE, we can exchange the fused $R$-matrix of the composite particle with the braiding operator, thereby giving rise to various $RBB$ relations.

\subsubsection{Fusion operators}

\begin{figure}[h]
 \begin{center}
   \includegraphics[width=0.75\linewidth]{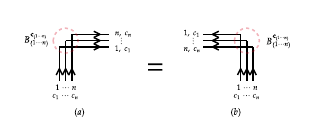}
 \end{center}
\caption{Two equivalent graphical representations of the fusion operator.  \label{fig:fo}}
\end{figure}

To establish connections with irreducible representations of $\mathfrak{gl}(N)$,  we consider a special braiding operator — the \textit{fusion operator} — 
defined as follows: 
\begin{equation}\label{fo1}
\begin{split}
B_{(1\cdots n)}^{\mb{c}_{(1\cdots n)}}=&\left[R^{c_1c_2}_{12}(0)\cdots R^{c_1c_n}_{1n}(0)\right]\left[R^{c_2c_3}_{23}(0)\cdots R^{c_2c_n}_{2n}(0)\right]\cdots \left[R^{c_{n-2},c_{n-1}}_{n-2,n-1}(0)R^{c_{n-1},c_n}_{n-2,n}(0)\right]R^{c_{n-1},c_n}_{n-1,n}(0)\\
=&\overset{\longrightarrow}{\prod_{1\leq i<j \leq n}} R_{ij}^{c_i,c_j}(0),
\end{split}
\end{equation}
where the symbol $\overset{\rightarrow}{\prod}_{i<j}$ indicates the product is taken in the lexicographical ordering on the pairs $(i,j)$. It is also graphically shown in Fig.~\ref{fig:fo} (a).

The fusion operator has an alternative expression, which is graphically represented by Fig.~\ref{fig:fo} (b):
\begin{equation}\label{fo2}
B_{(1\cdots n)}^{\mathbf{c}_{(1\cdots n)}}=\overset{\longleftarrow}{\prod_{1\leq i<j \leq n}} R_{ij}^{c_i,c_j}(0),
\end{equation}
where the factors $R_{ij}^{c_ic_j}(0)$ are taken in the reverse lexicographical ordering. An rigorous algebraic proof showing the equivalence between (\ref{fo1}) and (\ref{fo2}) can be readily obtained by mathematical induction (see \cite{d'Andecy:2017}). However, the equivalence is evident from Fig.~\ref{fig:fo}: By sequentially moving each particle line, starting from the $n$-th to the first in Fig.~\ref{fig:fo}(a), to the leftmost position through the application of the YBE, we can transform the configuration into Fig.~\ref{fig:fo}(b).

\begin{figure}[h]
 \begin{center}
   \includegraphics[width=0.8\linewidth]{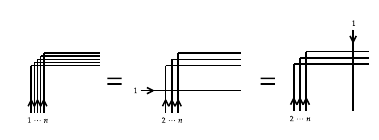}
 \end{center}
\caption{ First approach to decomposing the fusion operator. \label{fig:fodec1}}
\end{figure}

\begin{figure}[h]
 \begin{center}
   \includegraphics[width=0.8\linewidth]{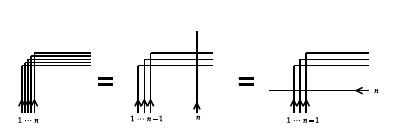}
 \end{center}
\caption{ Second approach to decomposing the fusion operator. \label{fig:fodec2}}
\end{figure}

The fusion operator $B_{(1\cdots n)}^{\mathbf{c}_{(1\cdots n)}}$ can also be decomposed in the following two ways:
\begin{eqnarray}\label{fodec1}
&&B_{(1\cdots n)}^{\mathbf{c}_{(1\cdots n)}}
=R_{1,(2\cdots n)}^{c_1,\mathbf{c}_{(2\cdots n)}}(0)B_{(2\cdots n)}^{\mathbf{c}_{(2\cdots n)}}
=B_{(2\cdots n)}^{\mathbf{c}_{(2\cdots n)}}R_{1,(n\cdots 2)}^{c_1,\mathbf{c}_{(n\cdots 2)}}(0),\\ 
\label{fodec2}
&&B_{(1\cdots n)}^{\mathbf{c}_{(1\cdots n)}}
=B_{(1\cdots n-1)}^{\mathbf{c}_{(1\cdots n-1)}}R_{(n-1\cdots 1),n}^{\mathbf{c}_{(n-1\cdots 1)},c_n}(0)
=R_{(1\cdots n-1),n}^{\mathbf{c}_{(1\cdots n-1)},c_n}(0)B_{(1\cdots n-1)}^{\mathbf{c}_{(1\cdots n-1)}},
\end{eqnarray}
which are shown in Fig.~\ref{fig:fodec1} and Fig.~\ref{fig:fodec2}.

Finally, we note that for the Yang solution of $R$-matrix (\ref{yangsl}), the fusion operator satisfies the relation:
\begin{equation}\label{Bpm}
B_{(1\cdots n)}^{\mathbf{c}_{(1\cdots n)}}=B_{(n\cdots 1)}^{-\mathbf{c}_{(n\cdots 1)}},
\end{equation}
which can be proved by induction:
Since $R_{ij}^{c_i,c_j}(0)=R_{ji}^{-c_j,-c_i}(0)$, it clearly holds for $n=2$. Suppose the relation holds for $n-1$. Then for the case $n$, due to (\ref{fodec2})  we have:
\begin{equation}
\begin{split}
B_{(1\cdots n)}^{\mathbf{c}_{(1\cdots n)}}=&R_{(1\cdots n-1),n}^{\mathbf{c}_{(1\cdots n-1)},c_n}(0)B_{(1\cdots n-1)}^{\mathbf{c}_{(1\cdots n-1)}}
=\left[R_{n-1,n}^{c_{n-1},c_n}(0)\cdots R_{1,n}^{c_1,c_n}(0)\right]B_{(n-1\cdots 1)}^{-\mathbf{c}_{(n-1\cdots 1)}}\\
=&\left[R_{n,n-1}^{-c_n,-c_{n-1}}(0)\cdots R_{n,1}^{-c_n,-c_1}(0)\right]B_{(n-1\cdots 1)}^{-\mathbf{c}_{(n-1\cdots 1)}}=B_{(n\cdots 1)}^{-\mathbf{c}_{(n\cdots 1)}}.
\end{split}
\end{equation}
\subsubsection{Invariant subspaces for irreducible representations of $\mathfrak{gl}(N)$}\label{Sec:fusion thm}
We now investigate the invariant subspace of the fused $R$-matrix $R_{(1\cdots n),(\ud{1}\cdots\ud{n}')}^{\mb{c}_{(1\cdots n)},\mb{c}_{(\ud{1}\cdots\ud{n}')}}(u)$. Define the subspaces:
\begin{equation}\label{tsb}
W=B_{(1\cdots n)}^{\mb{c}_{(1\cdots n)}}V^{\otimes n},\quad\quad \overline{W}=B_{(\ud{1}\cdots\ud{n}')}^{\mb{c}_{(\ud{1}\cdots\ud{n}')}}V^{\otimes n'}.
\end{equation}
From the $RBB$ relation (\ref{RBB}), in which the general braiding operator is chosen to be the fusion operator, it follows that  $W\otimes \overline{W}$ is invariant under the fused $R$-matrix:
\begin{equation}\label{eq5}
R_{(1\cdots n),(\ud{1}\cdots\ud{n}')}^{\mb{c}_{(1\cdots n)},\mb{c}_{(\ud{1}\cdots\ud{n}')}}(u)\in {\rm{End}}(W\otimes\overline{W}).
\end{equation}
More importantly, as will be detailed below, for specific chosen inhomogeneity parameters $\{c_i\}$ , the fusion operator $B_{(1\cdots n)}^{\mb{c}_{(1\cdots n)}}$ coincides with the primitive idempotent of the symmetric group $S_n$. Hence, according to the Schur-Weyl duality\cite{Schur-Weyl:1939}, $W$ becomes the irreducible representation space of $\mathfrak{gl}(N)$ and thus we obtain the reduced form of the fused $R$-matrix $R_{W\overline{W}}(u)$ by the restriction:
\begin{equation}\label{resR}
R_{W\overline{W}}(u):= R_{(1\cdots n),(\ud{1}\cdots\ud{n}')}^{\mb{c}_{(1\cdots n)},\mb{c}_{(\ud{1}\cdots\ud{n}')}}(u)\bigg|_{W\otimes\overline{W}}.
\end{equation}
The procedure described above is commonly referred to as the fusion procedure. A key step involves establishing a connection between the irreducible basis of the $\mathbb{C}[S_n]$-module (diagonal matrix unit) and the products of Yang $R$-matrices with specific parameters (the fusion operator), which can be traced back to the pioneering work of Jucys\cite{Jucys:1966}. The application of the fusion procedure in integrable systems originates from Cherendnik's work on constructing special bases for irreducible representations of Hecke algebras \cite{Cherednik:1989}. A rigorous proof was first given by Nazarov \cite{Nazarov:1998} and a slightly different version was proposed by Molev \cite{Molev:2008}. 

Now we give the details of the fusion theorem (we use a formulation as in \cite{d'Andecy:2017}): Let $\lambda$ be a partition of $n$ and let $\mathcal{T}$ be a standard Young tableau of shape $\lambda$. For a cell occupied by the number $k$ in the $i$-th row  and the $j$-th column  of $\mathcal{T}$, we define the content of this cell as $c(\mathcal{T}|k):=j-i$. We denote by $h(\lambda)$ the product of the hook lengths of the Young diagram $\lambda$. Then, the following evaluation of the fusion operator
\begin{equation}
E_{\mathcal{T}}:=\frac{1}{h(\lambda)}B_{(1\cdots n)}^{\mb{c}_{(1\cdots n)}}\bigg|_{c_i=c(\mathcal{T}|i),\,i=1,2,\cdots,n}
\end{equation}
gives: \begin{itemize}
         \item $E_{\mathcal{T}}$ is a primitive idempotent of $\mathbb{C}[S_n]$-module that generates a minimal left ideal $\mathbb{C}[S_n]E_{\mathcal{T}}$ .
         \item The subspace  $ W=E_{\mathcal{T}}V^{\otimes n}$ is an irreducible $\mathfrak{gl}(N)$-module.
       \end{itemize}
When the fused $R$-matrices are expressed in the irreducible representation subspaces of $\mathfrak{gl}(N)$ as given in (\ref{resR}), the fused YBE (\ref{fYBE}), which is now seen as an equation on the invariant tensor space $W\otimes \overline{W}\otimes \overline{\overline{W}}$,  reduces to:
\begin{equation}
R_{W\overline{W}}(u-v)R_{W\overline{\overline{W}}}(u)R_{\overline{W}\,\overline{\overline{W}}}(v)=R_{\overline{W}\,\overline{\overline{W}}}(v)
R_{W\overline{\overline{W}}}(u)R_{W\overline{W}}(u-v),
\end{equation}
where besides the subspaces $W$ and $\overline{W}$ defined in (\ref{tsb}), we also define: $\overline{\overline{W}}=B_{(\ud{\ud{1}}\cdots\ud{\ud{n}}'')}^{\mb{c}_{(\ud{\ud{1}}\cdots\ud{\ud{n}}'')}}V^{\otimes n''}$.

Remark: In the literature, it is also conventional to express the reduced fused $R$-matrix in the original tensor product space by  the projection operators (given by the primitive idempotent here). In this form, the reduced fused $R$-matrix is formulated as:
\begin{equation}\label{pfr}
E_{\mathcal{T}}E_{\mathcal{T}'}R_{(1\cdots n),(\ud{1}\cdots\ud{n}')}^{\mb{c}_{(1\cdots n)},\mb{c}_{(\ud{1}\cdots\ud{n}')}}(u)E_{\mathcal{T}}E_{\mathcal{T}'}
=R_{(1\cdots n),(\ud{1}\cdots\ud{n}')}^{\mb{c}_{(1\cdots n)},\mb{c}_{(\ud{1}\cdots\ud{n}')}}(u)E_{\mathcal{T}}E_{\mathcal{T}'},
\end{equation}
where $E_{\mathcal{T}}$ and $E_{\mathcal{T}'}$ project $V^{\otimes n}$ and $V^{\otimes n'}$ onto their irreducible subspaces $W$ and $\overline{W}$, respectively. We note that, in general, the matrix representation of (\ref{pfr})  contains many null rows and columns, and that upon their removal, the expression will reduce to the form of the restricted fused $R$-matrix in (\ref{resR}).

\begin{example}\label{ex1}
To illustrate the above fusion procedure, we examine the case $n=n'=2$ in (\ref{frm}) and consider the fused $R$-matrix $R_{(12),(\ud{1}\ud{2})}^{c_1,c_2,c_{\ud{1}},c_{\ud{2}}}(u)$. Following the construction in (\ref{frmxp}), for general inhomogeneities $c_1,c_2,c_{\ud{1}}$ and $c_{\ud{2}}$, the fused $R$-matrix is given by:
\begin{equation}
R_{(12),(\ud{1}\ud{2})}^{c_1,c_2,c_{\ud{1}},c_{\ud{2}}}(u)=R_{2\ud{1}}(c_2-c_{\ud{1}})R_{2\ud{2}}(c_2-c_{\ud{2}})R_{1\ud{1}}(c_1-c_{\ud{1}})R_{1\ud{2}}(c_1-c_{\ud{2}}),
\end{equation}
and two fusion operators, according to (\ref{fo1}), are
\begin{equation}
\begin{split}
B_{(12)}^{c_1c_2}=R_{12}(c_1-c_2),\quad B_{(\ud{1}\ud{2})}^{c_{\ud{1}}c_{\ud{2}}}=R_{\ud{1}\ud{2}}(c_{\ud{1}}-c_{\ud{2}}),
\end{split}
\end{equation}
together they satisfy the intertwining relation
\begin{equation}
R_{(12),(\ud{1}\ud{2})}^{c_1,c_2,c_{\ud{1}},c_{\ud{2}}}(u-v) B_{(12)}^{c_1c_2} B_{(\ud{1}\ud{2})}^{c_{\ud{1}}c_{\ud{2}}}
=B_{(12)}^{c_1c_2} B_{(\ud{1}\ud{2})}^{c_{\ud{1}}c_{\ud{2}}} R_{(21),(\ud{2}\ud{1})}^{c_2,c_1,c_{\ud{2}},c_{\ud{1}}}(u-v),
\end{equation}
which is graphically shown in Fig.~\ref{fig:rbb2}.

\begin{figure}[h]
 \begin{center}
   \includegraphics[width=0.8\linewidth]{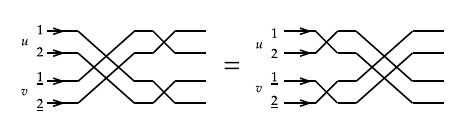}
 \end{center}
\caption{ $RBB$-relation for $n=n'=2$. \label{fig:rbb2}}
\end{figure}

We now consider the decomposition of tensor space $V_1\otimes V_2$. For the following two Young diagrams, the related fusion operators are:
\begin{equation}
\yng(2): \quad c_1=0, c_2=1,\quad B_{(12)}^{0,1}=R_{12}(-1)=\mathbb{I}_{12}+\mathbb{P}_{12},
\end{equation}
\begin{equation}
\yng(1,1):  \quad c_1=0, c_2=-1,\quad B_{(12)}^{0,-1}=R_{12}(1)=\mathbb{I}_{12}-\mathbb{P}_{12},
\end{equation}
which are indeed proportional to the projectors onto the symmetric and antisymmetric subspaces of $V^{\otimes 2}$, respectively. Similarly, for $V_{\ud{1}}\otimes V_{\ud{2}}$, two fusion operators are  $B^{0,1}_{(\ud{1}\ud{2})}$ and $B^{0,-1}_{(\ud{1}\ud{2})}$. If we further consider the case $\rm{dim}V=2$ and let $\{e_1,e_2\}$ be the standard basis of $V$, then we have:
\begin{equation}
\begin{split}
W:=&B_{(12)}^{0,1} V_1\otimes V_2={\rm{span}}\{e_1\otimes e_1,\,e_1\otimes e_2+e_2\otimes e_1,\,e_2\otimes e_2\},\\
\ud{W}:=&B^{0,1}_{(\ud{1}\ud{2})} V_{\ud{1}}\otimes V_{\ud{2}}={\rm{span}}\{e_1\otimes e_1,\,e_1\otimes e_2+e_2\otimes e_1,\,e_2\otimes e_2\},\\
W^{\perp}:=&B_{(12)}^{0,-1} V_1\otimes V_2={\rm{span}}\{e_1\otimes e_2-e_2\otimes e_1\},\\
\ud{W}^{\perp}:=&B^{0,-1}_{(\ud{1}\ud{2})} V_{\ud{1}}\otimes V_{\ud{2}}={\rm{span}}\{e_1\otimes e_2-e_2\otimes e_1\},
\end{split}
\end{equation}
and  $V_1\otimes V_2=W\oplus W^{\perp}$,$V_{\ud{1}}\otimes V_{\ud{2}}=\ud{W}\oplus \ud{W}^{\perp}$. We can calculate the reduced form of the fused $R$-matrix on these invariant subspaces, for instance,
\begin{equation}
R_{W,\ud{W}^{\perp}}(u):=R_{(12),(\ud{1}\ud{2})}^{(0,1),(0,-1)}(u)\bigg|_{{W\otimes\ud{W}^{\perp}}}.
\end{equation} 
The basis of $W\otimes\ud{W}^{\perp}$ is chosen to be:
\begin{equation}
\begin{split}
v_1&=2e_1\otimes e_1\otimes (e_1\otimes e_2-e_2\otimes e_1),\\
v_2&=(e_1\otimes e_2+e_2\otimes e_1)\otimes (e_1\otimes e_2-e_2\otimes e_1),\\
v_3&=2e_2\otimes e_2\otimes (e_1\otimes e_2-e_2\otimes e_1).
\end{split}
\end{equation}
The fused $R$-matrix acts on these basis vectors diagonally as:
\begin{equation}
\begin{split}
R_{(12),(\ud{1}\ud{2})}^{(0,1),(0,-1)}(u)v_i=\frac{u-1}{u+1}v_i,\quad i=1,2,3,
\end{split}
\end{equation}
thus, the reduced fused $R$-matrix turns out to be proportional to the identity matrix:
\begin{equation}
R_{W,\ud{W}^{\perp}}(u)=\frac{u-1}{u+1}\mathbb{I}_{3\times 3}.
\end{equation}
\end{example}

\subsection{Fused transfer matrices for closed spin chains} 

We first recall the fundamental monodromy matrix defined on the tensor product of a single auxiliary space $V_i$ and $n''$ quantum spaces $V_{(\ud{\ud{1}}\cdots\ud{\ud{n}}'')}=V_{\ud{\ud{1}}}\otimes\cdots\otimes V_{\ud{\ud{n}}''}$: 
\begin{equation}
T_{i;(\ud{\ud{1}}\cdots\ud{\ud{n}}'')}(u):=R_{i,\ud{\ud{1}}}^{c_i,c_{\ud{\ud{1}}}}(u)R_{i,\ud{\ud{2}}}^{c_i,c_{\ud{\ud{2}}}}(u)\cdots R_{i,\ud{\ud{n}}''}^{c_i,c_{\ud{\ud{n}}}''}(u),
\end{equation}
where the quantum spaces are typically suppressed, allowing us to write $T_i(u) \equiv T_{i;(\ud{\ud{1}}\cdots\ud{\ud{n}}'')}(u)$. Now we introduce $n$ monodromy matrices $\{T_i(u),\,i=1,2,\cdots, n\}$ and combine their auxiliary spaces to obtain the fused monodromy matrix as follows:
\begin{equation}
T_{(1\cdots n)}(u):=T_n(u)T_{n-1}(u)\cdots T_1(u).
\end{equation}
Note that only the auxiliary spaces of these fundamental monodromy matrices are fused, while their quantum spaces remain unchanged. Thus, by construction, we find the fused monodromy matrix is essentially the fused $R$-matrix:
\begin{equation}
T_{(1\cdots n)}(u)\equiv R_{(1\cdots n),(\ud{\ud{1}}\cdots\ud{\ud{n}}'')}^{\mb{c}_{(1\cdots n)},\mb{c}_{(\ud{\ud{1}}\cdots\ud{\ud{n}}'')}}(u).
\end{equation}
Due to the fused YBE (\ref{fYBE}), we obtain the fused $RTT$-relation:
\begin{equation}\label{fRTT}
R_{(1\cdots n),(\ud{1}\cdots \ud{n}')}^{\mathbf{c}_{(1\cdots n)},\mathbf{c}_{(\ud{1}\cdots\ud{n}')}}(u-v)T_{(1\cdots n)}(u)T_{(\ud{1}\cdots\ud{n}')}(v)
=T_{(\ud{1}\cdots\ud{n}')}(v)T_{(1\cdots n)}(u)R_{(1\cdots n),(\ud{1}\cdots \ud{n}')}^{\mathbf{c}_{(1\cdots n)},\mathbf{c}_{(\ud{1}\cdots\ud{n}')}}(u-v).
\end{equation}
Then the fused transfer matrix is obtained by tracing over the fused auxiliary space $V_{(1\cdots n)}=V_1\otimes\cdots\otimes V_n$:
\begin{equation}
{\tau}^{\mb{c}_{(1\cdots n)}}(u)=
\tr_{(1\cdots n)} T_{(1\cdots n)}(u)=\tr_{(1\cdots n)}R_{(1\cdots n),(\ud{\ud{1}}\cdots\ud{\ud{n}}'')}^{\mb{c}_{(1\cdots n)},\mb{c}_{(\ud{\ud{1}}\cdots\ud{\ud{n}}'')}}(u).
\end{equation}
Due to the fused $RTT$-relation (\ref{fRTT}), the fused transfer matrices with different spectral parameters and inhomogeneities form a commuting family:
\begin{equation}
\left[\tau(u)^{\mb{c}_{(1\cdots n)}},\tau(v)^{\mb{c}_{(\ud{1}\cdots\ud{n}')}}\right]=0,\quad \forall u,v,c_1 \cdots c_n,c_{\ud{1}}\cdots c_{\ud{n}'} \in \mathbb{C}.
\end{equation}
Clearly, the $n$-fused transfer matrix can be factorized into the multiplications of $n$ fundamental transfer matrices:
\begin{equation}
{\tau}^{\mb{c}_{(1\cdots n)}}(u)
=\tau^{c_n}(u)\tau^{c_{n-1}}(u)\cdots \tau^{c_1}(u).
\end{equation}
For a suitably chosen fusion operator $B_{(1\cdots n)}^{\mb{c}_{(1\cdots n)}}$, we have the auxiliary subspace $W:=B_{(1\cdots n)}^{\mb{c}_{(1\cdots n)}} V_{(1\cdots n)}$. The fused transfer matrix restricted to $W$ is 
\begin{equation}
\tau^{W}(u)=\tr_W T_{(1\cdots n)}(u)B_{(1\cdots n)}^{\mb{c}_{(1\cdots n)}}.
\end{equation}
The evaluation of such restricted transfer matrix $\tau^{W}(u)$ will be discussed in the next section.

\begin{example}\label{ex2}
Suppose we have three fundamental monodromy matrices, $T_1$, $T_2$, and $T_3$, each of which acts on the same quantum space $V_{\ud{1}} \otimes V_{\ud{2}}$. The fused monodromy matrix is then given by:
\begin{equation}
T_{(123)}(u)=T_3(u)T_2(u)T_1(u)=R_{3,(\ud{1}\ud{2})}^{c_3,c_{\ud{1}},c_{\ud{2}}}(u)R_{2,(\ud{1}\ud{2})}^{c_2,c_{\ud{1}},c_{\ud{2}}}(u)R_{1,(\ud{1}\ud{2})}^{c_1,c_{\ud{1}},c_{\ud{2}}}(u)
=R_{(123),(\ud{1}\ud{2})}^{\mb{c}_{(123)},\mb{c}_{(\ud{1}\ud{2})}}(u),
\end{equation}
which is also the fused $R$-matrix on $V_{(123)}\otimes V_{(\ud{1}\ud{2})}$. Taking the trace over the total auxiliary space $V_{(123)}$ yields the fused transfer matrix:
\begin{equation}
\tau^{c_1c_2c_3}(u)=\tr_{V_{(123)}}T_{(123)}(u).
\end{equation}
If we further consider the auxiliary subspace of $V_{(123)}$ constructed using the fusion operator $B^{c_1,c_2,c_3}_{(123)}$:
\begin{equation}
W=B^{c_1,c_2,c_3}_{(123)} V_{(123)},
\end{equation}
then the restricted fused monodromy matrix, that is the fused $R$-matrix restricted to the invariant subspace $W$, is given by:
\begin{equation}
T_W(u)=B^{c_1,c_2,c_3}_{(123)} R_{(123),(\ud{1}\ud{2})}^{\mb{c}_{(123)},\mb{c}_{(\ud{1}\ud{2})}}(u) B^{c_1,c_2,c_3}_{(123)},
\end{equation}
from which we obtain the restricted fused transfer matrix:
\begin{equation}
\tau^W(u)=\tr_W T_W(u).
\end{equation}
\end{example}

\section{Fusion procedure for open spin chains}
\begin{figure}[h]
 \begin{center}
   \includegraphics[width=0.8\linewidth]{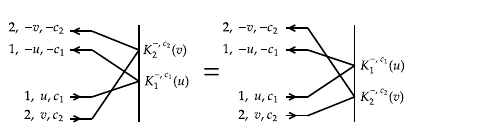}
 \end{center}
\caption{ Graphical representation of the reflection equation for $K^{-}(u)$ matrix with inhomogeneity parameters. \label{fig:re}}
\end{figure}

\begin{figure}[h]
 \begin{center}
   \includegraphics[width=0.8\linewidth]{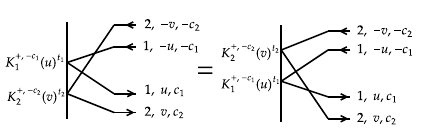}
 \end{center}
\caption{ Graphical representation of the dual reflection equation for $K^{+}(u)^t$ matrix with inhomogeneity parameters. \label{fig:dre}}
\end{figure}

In this section, we begin to investigate the general fusion procedure for open spin chains. The construction of integrable open spin chains was first introduced by Sklyanin in his seminal work \cite{Sklyanin:1988yz}. A key feature of this approach is the introduction of boundary reflection matrices $K^-(u)$ and $K^+(u)$,  which respectively generate the right and left boundary terms of the open spin chain Hamiltonian. To ensure integrability, the two reflection matrices $K^-(u)$ and $K^+(u)$ must satisfy the reflection equation (RE) and dual reflection equation (dual RE) respectively. In the general $\mathfrak{gl}(N)$ case, these equations read:
\begin{equation}\label{re}
R_{12}^{c_1c_2}(u-v)K_1^{-,c_1}(u)R_{21}^{c_2,-c_1}(u+v)K_2^{-,c_2}(v)=K_2^{-,c_2}(v)R_{12}^{c_1,-c_2}(u+v)K_1^{-,c_1}(u)R_{21}^{-c_2,-c_1}(u-v),
\end{equation}
and\footnote{Note that following the convention in \cite{Sklyanin:1988yz}, Eq.~(\ref{dre}) actually describes the dual RE for $K^{+}(u)^t$. By taking the full transpose in $V_1\otimes V_2$, we can readily obtain the dual RE for $K^+(u)$. We will elaborate on this point in the fusion case below.}
\begin{equation}\label{dre}
\begin{split}
&R_{12}^{-c_1,-c_2}(-u+v)K^{+,-c_1}_1(u)^{t_1}R_{21}^{-c_2,c_1}(-u-v+N)K_2^{+,-c_2}(v)^{t_2}\\
=&K_2^{+,-c_2}(v)^{t_2}R_{12}^{-c_1,c_2}(-u-v+N)K^{+,-c_1}_1(u)^{t_1}R_{21}^{c_2,c_1}(v-u),
\end{split}
\end{equation}
where $K^{-,c_1}_1(u)\equiv K^-_1(u+c_1),\,K^{+,-c_1}_1(u)^{t_1}\equiv K^{+}_1(u+c_1)^{t_1}$.  
The graphical representation for RE (\ref{re}) and dual RE (\ref{dre}) are shown in Fig.~\ref{fig:re} and Fig.~\ref{fig:dre}.


Based on the fundamental $K^-(u)$ and $K^+(u)$ reflection matrices, we now turn to the construction of the fused $K^-(u)$ matrices and the fused RE, as well as the fused $K^+(u)$ matrices and the fused dual RE. A concrete fusion procedure for open spin chains has been presented in \cite{Mezincescu:1991ke}, though restricted to the fusion of two auxiliary spaces. In the subsequent work \cite{Zhou:1995zy}, a $\mathfrak{su}(2)$ fusion rule for the models with general boundary conditions has been proposed, utilizing diagrammatic representations to elucidate fusion procedures. Building on these works, we will consider a general fusion procedure for open $\mathfrak{gl}(N)$ spin chains in arbitrary irreducible representations by specifically chosen fusion operators.

\subsection{Fused $K^{-}$-matrices and fused reflection equation}
\begin{figure}[h]
 \begin{center}
   \includegraphics[width=0.6\linewidth]{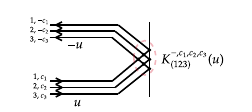}
 \end{center}
\caption{Graphical representation of the fused reflection matrix $K^{-,\mb{c}_{(123)}}_{(123)}(u)$.\label{fig:fK3}}
\end{figure}

The fused $K^-(u)$ matrix can be physically considered as the reflection matrix for the composite particles at the right boundary, where the in-coming  particles share a common spectral parameter $u$ but can carry distinct inhomogeneity parameters $c_i$. We will denote the fused $K^-(u)$ matrix for $n$-composite particles by $K^{-,\mb{c}_{(1\cdots n)}}_{(1\cdots n)}(u)$, with its graphical representation for $n=3$ given in Fig.~\ref{fig:fK3}.

\begin{figure}[h]
 \begin{center}
   \includegraphics[width=1.0\linewidth]{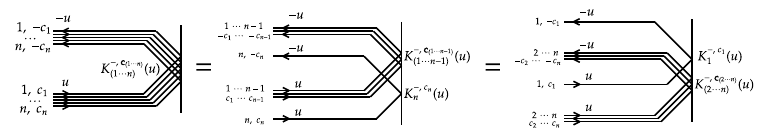}
 \end{center}
\caption{ Decompositions of the fused $K^{-}(u)$ matrix. \label{fig:dkm}}
\end{figure}

The exact expression for $K^{-,\mb{c}_{(1\cdots n)}}_{(1\cdots n)}(u)$ can be read off directly from the graph by associating suitable fundamental $R$-matrix or $K^-$-matrix with each graphical crossing point. For example, the $n=3$ gives:
\begin{equation}
K_{(123)}^{-,c_1,c_2,c_3}(u)=K_3^{-,c_3}(u)R_{23}^{c_2,-c_3}(2u)R_{13}^{c_1,-c_3}(2u)K_2^{-,c_2}(u)R_{12}^{c_1,-c_2}(2u)K^{c_1}_1(u).
\end{equation}
For general $n$, it follows recursively from the following two decomposition relations:
\begin{eqnarray}
K^{-,\mb{c}_{(1\cdots n)}}_{(1\cdots n)}(u)\label{dKm1}
&=&K^{-,c_n}_n(u)R_{(1\cdots n-1),n}^{\mb{c}_{(1\cdots n-1)},-c_n}(2u)K^{-,\mb{c}_{(1\cdots n-1)}}_{(1\cdots n-1)}(u)\\ \label{dKm2}
&=&K^{-,\mb{c}_{(2\cdots n)}}_{(2\cdots n)}(u)R_{1,(n\cdots 2)}^{c_1,-\mb{c}_{(n\cdots 2)}}(2u)K^{-,c_1}_1(u),
\end{eqnarray}
which correspond two processes displayed in Fig.~\ref{fig:dkm}.

Given the fused $R$-matrices and  $K^-$-matrices, the fused reflection equation can be read off from the reflection diagram of composite particles at the right boundary in Fig.~\ref{fig:fre}, yielding:
\begin{figure}[h]
 \begin{center}
   \includegraphics[width=0.6\linewidth]{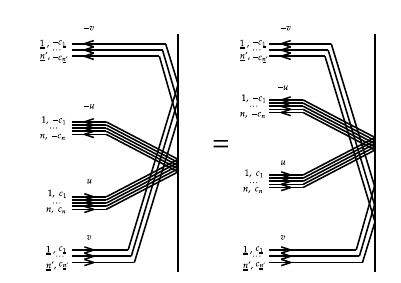}
 \end{center}
\caption{ Graphical representation of the reflection equation for fused $K^-(u)$-matrices. \label{fig:fre}}
\end{figure} 
\begin{equation}\label{fre}
\begin{split}
&R_{(1\cdots n),(\ud{1}\cdots \ud{n}')}^{\mathbf{c}_{(1\cdots n)},\mathbf{c}_{(\ud{1}\cdots\ud{n}')}}(u-v)
K^{-,\mathbf{c}_{(1\cdots n)}}_{(1\cdots n)}(u)
R_{(\ud{1}\cdots\ud{n}'),(n\cdots 1)}^{\mb{c}_{(\ud{1}\cdots\ud{n}')},-\mb{c}_{(n\cdots 1)}}(u+v)
K^{-,\mb{c}_{(\ud{1}\cdots\ud{n}')}}_{(\ud{1}\cdots\ud{n}')}(v)\\
=&K^{-,\mb{c}_{(\ud{1}\cdots\ud{n}')}}_{(\ud{1}\cdots\ud{n}')}(v)
R_{(1\cdots n),(\ud{n}'\cdots \ud{1})}^{\mathbf{c}_{(1\cdots n)},-\mathbf{c}_{(\ud{n}'\cdots\ud{1})}}(u+v)
K^{-,\mathbf{c}_{(1\cdots n)}}_{(1\cdots n)}(u)
R_{(\ud{n}'\cdots\ud{1}),(n\cdots 1)}^{-\mb{c}_{(\ud{n}'\cdots\ud{1})},-\mb{c}_{(n\cdots 1)}}(u-v).
\end{split}
\end{equation}
Based on the fundamental RE (\ref{re}) and the decomposition relations (\ref{dKm1}-\ref{dKm2}), a rigorous proof for the fused RE (\ref{fre}) can be straightforwardly obtained through mathematical induction. We also note that (\ref{fre}) represents the most generic form of the fused RE, which imposes no symmetry requirements on the fused $R$-matrices. If we further impose the  P-symmetry condition (\ref{psym}) on the $R$-matrices, namely:
\begin{equation}
\begin{split}
R_{(1\cdots n),(\ud{1}\cdots \ud{n}')}^{\mathbf{c}_{(1\cdots n)},\mathbf{c}_{(\ud{1}\cdots\ud{n}')}}(u-v)=R_{(\ud{n}'\cdots\ud{1}),(n\cdots 1)}^{-\mb{c}_{(\ud{n}'\cdots\ud{1})},-\mb{c}_{(n\cdots 1)}}(u-v),\\
R_{(\ud{1}\cdots\ud{n}'),(n\cdots 1)}^{\mb{c}_{(\ud{1}\cdots\ud{n}')},-\mb{c}_{(n\cdots 1)}}(u+v)=R_{(1\cdots n),(\ud{n}'\cdots \ud{1})}^{\mathbf{c}_{(1\cdots n)},-\mathbf{c}_{(\ud{n}'\cdots\ud{1})}}(u+v),
\end{split}
\end{equation}
then we observe that the right-hand side of the fused RE (\ref{fre}) is precisely the order-reversed expression of the left-hand side. 
\subsubsection{Invariant subspaces of the fused $K^{-}$-matrices}
To identify the invariant subspace of \(K^{-,\mathbf{c}_{(1\cdots n)}}_{(1\cdots n)}\) in \(V^{\otimes n}\), we must establish the intertwining relation between \(K^{-,\mathbf{c}_{(1\cdots n)}}_{(1\cdots n)}\) and the fusion operator \(B_{(1\cdots n)}^{\mathbf{c}_{(1\cdots n)}}\). Setting \(u = v\) in the fundamental RE (\ref{re}) yields:
\begin{equation}
R_{12}^{c_1c_2}(0)K_1^{-,c_1}(u)R_{21}^{c_2,-c_1}(2u)K_2^{-c_2}(u)=K^{-,c_2}_2(u)R_{12}^{c_1,-c_2}(2u)K^{-,c_1}_1(u)R_{21}^{-c_2,-c_1}(0),
\end{equation}
which constitutes the \(n=2\) intertwining relation:
\begin{equation}
K^{-,c_1c_2}_{(12)}(u)B_{(21)}^{-c_2,-c_1}=B_{(12)}^{c_1c_2}K_{(21)}^{-,c_2c_1}(u).
\end{equation}
For general $n$, we propose:
\begin{equation}\label{interKm}
K^{-,\mb{c}_{(1\cdots n)}}_{(1\cdots n)}(u) B_{(n\cdots 1)}^{-\mb{c}_{(n\cdots 1)}} =B_{(1\cdots n)}^{\mb{c}_{(1\cdots n)}} K_{(n\cdots 1)}^{-,\mb{c}_{(n\cdots 1)}}(u).
\end{equation}
The proof proceeds by mathematical induction. Assuming validity for $n-1$, the $n$-case derivation is:
\begin{equation}
\begin{split}
B_{(1\cdots n)}^{\mb{c}_{(1\cdots n)}}K^{-,\mb{c}_{(n\cdots 1)}}_{(n\cdots 1)}(u)
=&\left[B_{(1\cdots n-1)}^{\mb{c}_{(1\cdots n-1)}}R_{(n-1\cdots 1),n}^{\mb{c}_{(n-1,\cdots 1)},c_n}(0)\right]
\left[K^{-,\mb{c}_{(n-1\cdots 1)}}_{(n-1\cdots 1)}(u) R_{n,(1\cdots n-1)}^{c_n,-\mb{c}_{(1\cdots n-1)}}(2u)K^{-,c_n}_{n}(u)\right]\\
=&B_{(1\cdots n-1)}^{\mb{c}_{(1\cdots n-1)}}\left[K^{-,c_n}_n(u)R_{(n-1\cdots 1),n}^{\mb{c}_{(n-1\cdots 1)},-c_n}(2u)K^{-,\mb{c}_{(n-1\cdots 1)}}_{(n-1\cdots 1)}(u)R_{n,(1\cdots n-1)}^{-c_n,-\mb{c}_{(1\cdots n-1)}}(0)\right]\\
=&K^{-,c_n}_n(u)\left[R_{(1\cdots n-1),n}^{\mb{c}_{(1\cdots n-1)},-c_n}(2u)B_{(1\cdots n-1)}^{\mb{c}_{(1\cdots n-1)}}\right]K^{-,\mb{c}_{(n-1\cdots 1)}}_{(n-1\cdots 1)}(u)R_{n,(1\cdots n-1)}^{-c_n,-\mb{c}_{(1\cdots n-1)}}(0)\\
=&K^{-,c_n}_n(u)R_{(1\cdots n-1),n}^{\mb{c}_{(1\cdots n-1)},-c_n}(2u)\left[K^{-,\mb{c}_{(n-1\cdots 1)}}_{(1\cdots n-1)}(u)B_{(n-1\cdots 1)}^{-\mb{c}_{(n-1\cdots 1)}}\right]R_{n,(1\cdots n-1)}^{-c_n,-\mb{c}_{(1\cdots n-1)}}(0)\\
=&K^{-,\mb{c}_{(1\cdots n)}}_{(1\cdots n)}(u) B_{(n\cdots 1)}^{-\mb{c}_{(1\cdots n)}},
\end{split}
\end{equation}
where in the proof we have used the decomposition relations of the fusion operator (\ref{fodec1},\ref{fodec2}), the decomposition relations for $K^-$-matrix (\ref{dKm1}) and (\ref{dKm2}), and the fused  RE (\ref{fre}). Finally, combining the intertwining relation (\ref{interKm}) with the relation (\ref{Bpm}), we find the invariant subspace of $K^{-,\mb{c}_{(1\cdots n)}}_{(1\cdots n)}(u)$ is:
\begin{equation}\label{eq6}
K^{-,\mb{c}_{(1\cdots n)}}_{(1\cdots n)}(u)\in{\rm{End}}\left(B_{(1\cdots n)}^{\mb{c}_{(1\cdots n)}} V^{\otimes n}\right).
\end{equation}

\begin{example}\label{ex3}
We consider a simple 2-fused $K_{(12)}^{-,c_1c_2}(u)$ (corresponding to the fused $R$-matrix in \ref{ex1}), which, by construction, is given by:
\begin{equation}\label{exf1}
\begin{split}
K^{-,c_1c_2}_{(12)}(u)=&K^{-,c_2}_2(u)R_{12}^{c_1c_2}(2u)K^{-,c_1}_1(u)\\
=&K^{-}_1(u+c_1)K^-_2(u+c_2)-\frac{1}{2u+c_1-c_2}\mathbb{P}_{12}K^-_1(u+c_2)K^-_1(u+c_1).
\end{split}
\end{equation}
By the fusion operators $B_{(12)}^{0,1}$ and $B_{(12)}^{0,-1}$, we obtain the symmetric and antisymmetric invariant subspace, respectively:
$W=S^2V^{\otimes 2}:=B_{(12)}^{0,1}V^{\otimes 2},\,W^{\perp}=\Lambda^2V^{\otimes 2}:=B_{(12)}^{0,-1} V^{\otimes 2}$. Therefore, the restricted 2-fused $K^-$-matrices are:
\begin{equation}\label{exf2}
K^-_W= B_{(12)}^{0,1} K^{-,0,1}_{(12)}(u) B_{(12)}^{0,1},\quad K^-_{W^{\perp}}= B_{(12)}^{0,-1} K^{-,0,-1}_{(12)}(u) B_{(12)}^{0,-1}.
\end{equation} 
Given a concrete solution $K^-(u)$ of the fundamental RE (\ref{re}), substituting it into (\ref{exf1}) and (\ref{exf2}), we can readily obtain the explicit expressions of the fused $K^-$-matrix and its restricted forms. 
\end{example}

\subsection{Fused $K^{+}$-matrices and fused dual reflection equation}
\begin{figure}[h]
 \begin{center}
   \includegraphics[width=0.6\linewidth]{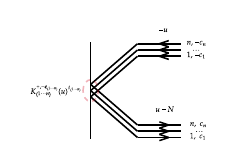}
 \end{center}
\caption{ Graphical representation of the fused reflection matrix $K^{+,-\mb{c}_{(1\cdots n)}}_{(1\cdots n)}(u)^{t_{(1\cdots n)}}.$ \label{fig:fKpn}}
\end{figure}

Analogous to the fused $K^-$-matrix, we now introduce the fused $K^+$-matrix as the scattering matrix for composite particles at the left boundary. Pictorially, $K^{+,-\mb{c}_{(1\cdots n)}}_{(1\cdots n)}(u)^{t_{(1\cdots n)}}$ is shown in Fig.~\ref{fig:fKpn} .

Note that after left-boundary reflection, the rapidity parameters (spectral and inhomogeneity parameters) not only reverse sign (as occurs at the right boundary) but also acquire an additional constant shift $N$  in the general $\mathfrak{gl}(N)$ case, as specified below:
\begin{eqnarray}
-(u+c_i)\rightarrow u+c_i-N\quad \Longleftrightarrow \quad 
\left\{
\begin{array}{l}
-u\rightarrow u-N,\\
-c_i\rightarrow c_i
\end{array}
\right.
\end{eqnarray}
We henceforth adopt the convention of putting the shift $N$ within the transformation of the spectral parameter $u$ under reflection: $-u\rightarrow u-N$.

The explicit form of $K^{+,-\mb{c}_{(1\cdots n)}}_{(1\cdots n)}(u)^{t_{(1\cdots n)}}$ can be directly extracted from  Fig.~\ref{fig:fKpn}. For $n=2$, it gives:
\begin{equation}\label{kp1}
K^{+,-c_1,-c_2}_{(12)}(u)^{t_1t_2}=K^{+,-c_2}_2(u)^{t_2}R_{12}^{-c_1,c_2}(-2u+N)K^{+,-c_1}_1(u)^{t_1}.
\end{equation}
For general $n$, it can be obtained recursively from the following decomposition relations:
\begin{eqnarray}
K^{+,-\mathbf{c}_{(1\cdots n)}}_{(1\cdots n)}(u)^{t_{(1\cdots n)}}
&=&K^{+,-\mathbf{c}_{(2\cdots n)}}_{(2\cdots n)}(u)^{t_{(2\cdots n)}}\label{dKp1}
R_{1,(n\cdots 2)}^{-c_1,\mathbf{c}_{(n\cdots 2)}}(-2u+N)K^{+,-c_1}_1(u)^{t_1}\\
&=&K^{+,-c_n}_n(u)^{t_n}R_{(1\cdots n-1),n}^{-\mathbf{c}_{(1\cdots n-1)},c_n}(-2u+N)K^{+,-\mathbf{c}_{(1\cdots n-1)}}_{(1\cdots n-1)}(u)^{t_{(1\cdots n-1)}}.\label{dKp2}
\end{eqnarray}
Graphically, these decompositions correspond to the following two processes:
\begin{figure}[h]
 \begin{center}
   \includegraphics[width=1.0\linewidth]{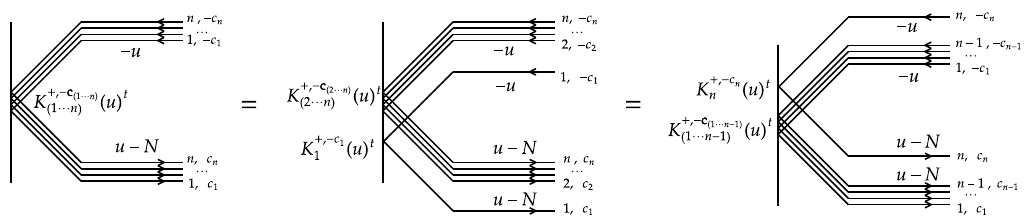}
 \end{center}
\caption{ Decompositions of the fused $K^{+}(u)$ matrix. \label{fig:decpKp}}
\end{figure} 

\begin{figure}[h]
 \begin{center}
   \includegraphics[width=0.85\linewidth]{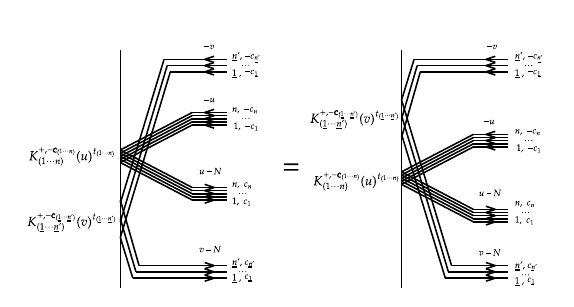}
 \end{center}
\caption{Graphical representation of the dual RE for $K^{+}(u)^t$-matrices. \label{fig:fdre}}
\end{figure}

Similar to the right boundary case, we establish two equivalent processes of the scattering and reflection for the composite particles at the left boundary , as illustrated in Fig.~\ref{fig:fdre}.

Then the dual reflection equation for the fused $K^+(u)^{t}$-matrices can be read off from Fig.~\ref{fig:fdre} as:
\begin{equation}\label{fdre}
  \begin{split}
  &R_{(1\cdots n),(\ud{1}\cdots \ud{n}')}^{-\mathbf{c}_{(1\cdots n)},-\mathbf{c}_{(\ud{1}\cdots \ud{n}')}}(-u+v) 
  K^{+,-\mathbf{c}_{(1\cdots n)}}_{(1\cdots n)}(u)^{t_{(1\cdots n)}} 
  R_{(\ud{1}\cdots \ud{n}'),(n\cdots 1)}^{-\mathbf{c}_{(\ud{1}\cdots\ud{n}')},\mathbf{c}_{(n\cdots 1)}}(-v-u+N) K^{+,-\mathbf{c}_{(\ud{1}\cdots \ud{n}')}}_{(\ud{1}\cdots\ud{n}')}(v)^{t_{(\ud{1}\cdots\ud{n}')}}\\
  =&K^{+,-\mathbf{c}_{(\ud{1}\cdots \ud{n}')}}_{(\ud{1}\cdots\ud{n}')}(v)^{t_{(\ud{1}\cdots\ud{n}')}}
  R_{(1\cdots n),(\ud{n}'\cdots\ud{1})}^{-\mathbf{c}_{(1\cdots n)},\mathbf{c}_{(\ud{n}'\cdots\ud{1})}}(-u-v+N)
  K^{+,-\mathbf{c}_{(1\cdots n)}}_{(1\cdots n)}(u)^{t_{(1\cdots n)}}
  R_{(\ud{n}'\cdots\ud{1}),(n\cdots 1)}^{\mathbf{c}_{(\ud{n}'\cdots\ud{1})},\mathbf{c}_{(n\cdots 1)}}(v-u).
  \end{split}
\end{equation}
Similarly, through mathematical induction, we can rigorously demonstrate that the fundamental dual RE (\ref{dre}) and decomposition relations (\ref{dKp1}-\ref{dKp2}) yield the fused RE (\ref{fdre}). Equivalently, by performing the total transpose in $V^{\otimes n}\otimes V^{\otimes n'}$, we obtain the dual reflection equation for the fused $K^+(u)$-matrices: 
\begin{equation}
\begin{split}
&R_{(n\cdots 1),(\ud{n}'\cdots\ud{1})}^{-\mb{c}_{(n\cdots 1)},-\mb{c}_{(\ud{n}'\cdots\ud{1})}}(-u+v)K^{+,-\mb{c}_{(1\cdots n)}}_{(1\cdots n)}(u)
R_{(\ud{n}'\cdots\ud{1}),(1\cdots n)}^{-\mb{c}_{(\ud{n}'\cdots\ud{1})},\mb{c}_{(1\cdots n)}}(-u-v+N)K^{+,-\mb{c}_{(\ud{1}\cdots\ud{n}')}}_{(\ud{1}\cdots\ud{n}')}(v)\\
=&K^{+,-\mb{c}_{(\ud{1}\cdots\ud{n}')}}_{(\ud{1}\cdots\ud{n}')}(v) R_{(n\cdots 1),(\ud{1}\cdots\ud{n}')}^{-\mb{c}_{(n\cdots 1)},\mb{c}_{(\ud{1}\cdots\ud{n}')}}(-v-u+N)
K^{+,-\mb{c}_{(1\cdots n)}}_{(1\cdots n)}(u)
R_{(\ud{1}\cdots\ud{n}'),(1\cdots n)}^{\mb{c}_{(\ud{1}\cdots\ud{n}')},\mb{c}_{(1\cdots n)}}(-u+v).
\end{split}
\end{equation}

\subsubsection{Invariant subspaces of the fused $K^{+}$-matrices}
Now we consider the invariant subspace of $K^{+,-\mathbf{c}_{(1\cdots n)}}_{(1\cdots n)}(u)$. In the fundamental dual RE, letting $u=v$, we obtain:
\begin{equation}
\begin{aligned}
R_{12}^{-c_1,-c_2}(0)K^{+,-c_1}_1(u)^{t_1}R_{21}^{-c_2,c_1}(-2u+N)K^{+,-c_2}_2(u)^{t_2}\\=
K^{+,-c_2}_2(u)^{t_2}R_{12}^{-c_1,c_2}(-2u+N)K^{+,-c_1}_1(u)^{t_1}R_{21}^{c_2,c_1}(0),
\end{aligned}
\end{equation}
which shows that
\begin{equation}
\begin{aligned}
B_{(12)}^{-c_1,-c_2}K^{+,-c_2,-c_1}_{(21)}(u)^{t_1t_2}=K^{+,-c_1,-c_2}_{(12)}(u)^{t_1t_2} B_{(21)}^{c_2,c_1}.
\end{aligned}
\end{equation}
In general, we would have the relation:
\begin{equation}\label{KBt}
\begin{aligned}
B_{(1\cdots n)}^{-\mathbf{c}_{(1\cdots n)}} K^{+,-\mathbf{c}_{(n\cdots 1)}}_{(n\cdots 1)}(u)^{t_{(1\cdots n)}}=K^{+,-\mathbf{c}_{(1\cdots n)}}_{(1\cdots n)}(u)^{t_{(1\cdots n)}} B_{(n\cdots 1)}^{\mathbf{c}_{(n\cdots 1)}},
\end{aligned}
\end{equation}
which can be proved by induction: Assume the relation holds for $n-1$, then for $n$, 
\begin{equation}
\begin{aligned}
&B_{(1\cdots n)}^{-\mathbf{c}_{(1\cdots n)}} K^{+,-\mathbf{c}_{(n\cdots 1)}}_{(n\cdots 1)}(u)^{t_{(1\cdots n)}}\\
=&\left[B_{(1\cdots n-1)}^{-\mathbf{c}_{(1\cdots n-1)}}R_{(n-1\cdots 1),n}^{-\mathbf{c}_{(n-1\cdots 1),-c_n}}(0)\right]
\left[K^{+,-\mathbf{c}_{(n-1\cdots 1)}}_{(n-1\cdots 1)}(u)^{t_{(1\cdots n-1)}} R_{n,(1\cdots n-1)}^{-c_n,\mathbf{c}_{(1\cdots n-1)}}(-2u+N) 
K^{+,-c_n}_n(u)^{t_n}\right]\\
=&B_{(1\cdots n-1)}^{-\mathbf{c}_{(1\cdots n-1)}}\left[K^{+,-c_n}_n(u)^{t_n}R_{(n-1\cdots 1),n}^{-\mathbf{c}_{(n-1\cdots 1),c_n}}(-2u+N)
K^{+,-\mathbf{c}_{(n-1\cdots 1)}}_{(n-1\cdots 1)}(u)^{t_{(1\cdots n-1)}}
R_{n,(1\cdots n-1)}^{c_n,\mathbf{c}_{(1\cdots n-1)}}(0)\right]\\
=&K^{+,-c_n}_n(u)^{t_n}\left[R_{(1\cdots n-1),n}^{-\mathbf{c}_{(1\cdots n-1)},c_n}(-2u+N)B_{(1\cdots n-1)}^{-\mathbf{c}_{(1\cdots n-1)}}\right]K^{+,-\mathbf{c}_{(n-1\cdots 1)}}_{(n-1\cdots 1)}(u)^{t_{(1\cdots n-1)}}
R_{n,(1\cdots n-1)}^{c_n,\mathbf{c}_{(1\cdots n-1)}}(0)\\
=&K^{+,-c_n}_n(u)^{t_n}R_{(1\cdots n-1),n}^{-\mathbf{c}_{(1\cdots n-1)},c_n}(-2u+N)
\left[K^{+,-\mathbf{c}_{(1\cdots n-1)}}_{(1\cdots n-1)}(u)^{t_{(1\cdots n-1)}}B_{(n-1\cdots 1)}^{-\mathbf{c}_{(n-1\cdots 1)}}\right]R_{n,(1\cdots n-1)}^{c_n,\mathbf{c}_{(1\cdots n-1)}}(0)\\
=&K^{+,-\mathbf{c}_{(1\cdots n)}}_{(1\cdots n)}(u)^{t_{(1\cdots n)}} B_{(n\cdots 1)}^{\mathbf{c}_{(n\cdots 1)}},
\end{aligned}
\end{equation}
where in the proof we have used the decomposition relations of the fusion operator (\ref{fodec1},\ref{fodec2}), the fused dual RE (\ref{fdre}) and the decomposition relations for $K^+$-matrix (\ref{dKp1}) and (\ref{dKp2}).

Now taking the total transpose in $V^{\otimes n}$ to both sides of the equation (\ref{KBt}), we obtain:
\begin{equation}\label{KBt2}
 K^{+,-\mathbf{c}_{(1\cdots n)}}_{(1\cdots n)}(u)\left[B_{(n\cdots 1)}^{-\mathbf{c}_{(n\cdots 1)}}\right]^{t_{(1\cdots n)}}= \left[B_{(1\cdots n)}^{\mathbf{c}_{(1\cdots n)}}\right]^{t_{(1\cdots n)}}K^{+,-\mathbf{c}_{(n\cdots 1)}}_{(n\cdots 1)}(u).
\end{equation}
Since the fundamental $R$-matrix (\ref{yangsl}) has the property: $[R_{ij}^{c_ic_j}(u)]^{t_it_j}=R^{c_ic_j}_{ij}(u)$, we find
\begin{equation}
\begin{split}
\left[B_{(1\cdots n)}^{\mathbf{c}_{(1\cdots n)}}\right]^{t_{(1\cdots n)}}&=\left[(R_{12}^{c_1c_2}(0)\cdots R_{1n}^{c_1c_n}(0))\cdots R_{n-1,n}^{c_{n-1}c_n}(0)\right]^{t_{(1\cdots n)}}\\
&=R_{n-1,n}^{c_{n-1}c_n}(0)^{t_{n-1}t_n}\cdots \left[R_{1n}^{c_1c_n}(0)^{t_1t_n}\cdots R_{12}^{c_1c_2}(0)^{t_1t_2}\right]\\
&=R_{n-1,n}^{c_{n-1}c_n}(0)\cdots \left[R_{1n}^{c_1c_n}(0)\cdots R_{12}^{c_1c_2}(0)\right]=\overset{\longleftarrow}{\prod_{1\leq i<j \leq n}} R_{ij}^{c_i,c_j}(0)=B_{(1\cdots n)}^{\mathbf{c}_{(1\cdots n)}}.
\end{split}
\end{equation}
Also remind of the relation (\ref{Bpm}), we thus conclude from (\ref{KBt2}) that $B_{(1\cdots n)}^{\mathbf{c}_{(1\cdots n)}}V^{\otimes n}$ is the invariant subspace of the reflection operator $K^{+,-\mathbf{c}_{(1\cdots n)}}_{(1\cdots n)}$:
\begin{equation}
  K^{+,-\mathbf{c}_{(1\cdots n)}}_{(1\cdots n)}\in {\rm{End}}\left(B_{(1\cdots n)}^{\mathbf{c}_{(1\cdots n)}}V^{\otimes n}\right).
\end{equation}
\subsection{Fused transfer matrices for open spin chains}
In this section, we will examine the construction of fused transfer matrices for open spin chains. We begin by introducing the fused double-row monodromy matrix and exploring the RE it satisfies. Next, we construct the fused transfer matrix using the double-row monodromy matrix and the fused $K^+$-matrix. We will then demonstrate the commutativity property of the fused transfer matrices. Finally, we evaluate the fused transfer matrices in both the total space and the invariant subspaces.
\subsubsection{Fused double-row monodromy matrices}
The fused double-row monodromy matrix $U_{(1\cdots n)}(u)$ is defined as:
\begin{equation}\label{fdrm}
U_{(1\cdots n)}(u)\equiv U_{(1\cdots n),(\ud{\ud{1}}\cdots \ud{\ud{n}}'')}(u)
=R_{(1\cdots n),(\ud{\ud{1}}\cdots\ud{\ud{n}}'')}^{\mathbf{c}_{(1\cdots n)},\mathbf{c}_{(\ud{\ud{1}}\cdots\ud{\ud{n}}'')}}(u)
K^{-,\mathbf{c}_{(1\cdots n)}}_{(1\cdots n)}(u)
R_{(\ud{\ud{1}}\cdots\ud{\ud{n}}''),(n\cdots 1)}^{\mathbf{c}_{(\ud{\ud{1}}\cdots\ud{\ud{n}}'')},-\mathbf{c}_{(n\cdots 1)}}(u),
\end{equation}
where $V^{\otimes n''}=V_{\ud{\ud{1}}}\otimes\cdots\otimes V_{\ud{\ud{n}}''}$ is the internal space.
Pictorially, the fused double-row monodromy matrix can be seen as a modified right boundary reflection matrix with inner degrees of freedom, which is shown below in Fig.~\ref{fig:fdrm}.
\begin{figure}[h]
 \begin{center}
   \includegraphics[width=0.85\linewidth]{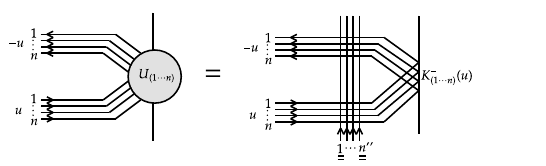}
 \end{center}
\caption{Fused double-row monodromy matrix as a modified reflection matrix. \label{fig:fdrm}}
\end{figure} 

\begin{figure}[h]
 \begin{center}
   \includegraphics[width=0.85\linewidth]{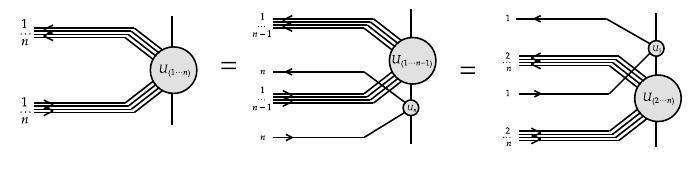}
 \end{center}
\caption{Decompositions of the fused double-row monodromy matrix. \label{fig:udc}}
\end{figure}

Similar to the decompositions of $K^-$-matrices (\ref{dKm1}) and (\ref{dKm2}), $U_{(1\cdots n)}$ can be decomposed as:
\begin{eqnarray}\label{udc1}
U_{(1\cdots n)}(u)&=&U_n(u) R_{(1\cdots n-1),n}^{\mb{c}_{(1\cdots n-1)},-c_{n}}(2u)U_{(1\cdots n-1)}(u)\\\label{udc2}
&=& U_{(2\cdots n)}(u) R_{1,(n\cdots 2)}^{c_1,-\mb{c}_{(n\cdots 2)}}(2u) U_1(u),
\end{eqnarray}
with the graphical representations given in Fig.~\ref{fig:udc}.

The fused double-row monodromy matrix $U_{(1\cdots n)}(u)$ obey the same fused RE (\ref{fre}) as demonstrated below:

\begin{equation}\label{drre}
\begin{split}
&R_{(1\cdots n),(\ud{1}\cdots\ud{n}')}^{\mb{c}_{(1\cdots n)},\mb{c}_{(\ud{1}\cdots\ud{n}')}}(u-v) U_{(1\cdots n)}(u)
R_{(\ud{1}\cdots\ud{n}'),(n\cdots 1)}^{\mb{c}_{(\ud{1}\cdots\ud{n}')},-\mb{c}_{(n\cdots 1)}}(u+v) U_{(\ud{1}\cdots\ud{n}')}(v)\\
=& R_{(1\cdots n),(\ud{1}\cdots\ud{n}')}^{\mb{c}_{(1\cdots n)},\mb{c}_{(\ud{1}\cdots\ud{n}')}}(u-v) 
\left[R_{(1\cdots n),(\ud{\ud{1}}\cdots\ud{\ud{n}}'')}^{\mathbf{c}_{(1\cdots n)},\mathbf{c}_{(\ud{\ud{1}}\cdots\ud{\ud{n}}'')}}(u)
K^{-,\mathbf{c}_{(1\cdots n)}}_{(1\cdots n)}(u)
R_{(\ud{\ud{1}}\cdots\ud{\ud{n}}''),(n\cdots 1)}^{\mathbf{c}_{(\ud{\ud{1}}\cdots\ud{\ud{n}}'')},-\mathbf{c}_{(n\cdots 1)}}(u)\right]R_{(\ud{1}\cdots\ud{n}'),(n\cdots 1)}^{\mb{c}_{(\ud{1}\cdots\ud{n}')},-\mb{c}_{(n\cdots 1)}}(u+v) \\
&
\left[R_{(\ud{1}\cdots \ud{n}'),(\ud{\ud{1}}\cdots\ud{\ud{n}}'')}^{\mathbf{c}_{(\ud{1}\cdots \ud{n}')},\mathbf{c}_{(\ud{\ud{1}}\cdots\ud{\ud{n}}'')}}(v)
K^{-,\mathbf{c}_{(\ud{1}\cdots \ud{n}')}}_{(\ud{1}\cdots \ud{n}')}(v)
R_{(\ud{\ud{1}}\cdots\ud{\ud{n}}''),(\ud{n}'\cdots \ud{1})}^{\mathbf{c}_{(\ud{\ud{1}}\cdots\ud{\ud{n}}'')},-\mathbf{c}_{(\ud{n}'\cdots \ud{1})}}(v)\right]\\
=&R_{(1\cdots n),(\ud{1}\cdots\ud{n}')}^{\mb{c}_{(1\cdots n)},\mb{c}_{(\ud{1}\cdots\ud{n}')}}(u-v)R_{(1\cdots n),(\ud{\ud{1}}\cdots\ud{\ud{n}}'')}^{\mathbf{c}_{(1\cdots n)},\mathbf{c}_{(\ud{\ud{1}}\cdots\ud{\ud{n}}'')}}(u)
K^{-,\mathbf{c}_{(1\cdots n)}}_{(1\cdots n)}(u)\\
&
\left[R_{(\ud{\ud{1}}\cdots\ud{\ud{n}}''),(n\cdots 1)}^{\mathbf{c}_{(\ud{\ud{1}}\cdots\ud{\ud{n}}'')},-\mathbf{c}_{(n\cdots 1)}}(u)
R_{(\ud{1}\cdots\ud{n}'),(n\cdots 1)}^{\mb{c}_{(\ud{1}\cdots\ud{n}')},-\mb{c}_{(n\cdots 1)}}(u+v) 
R_{(\ud{1}\cdots \ud{n}'),(\ud{\ud{1}}\cdots\ud{\ud{n}}'')}^{\mathbf{c}_{(\ud{1}\cdots \ud{n}')},\mathbf{c}_{(\ud{\ud{1}}\cdots\ud{\ud{n}}'')}}(v)
\right]K^{-,\mathbf{c}_{(\ud{1}\cdots \ud{n}')}}_{(\ud{1}\cdots \ud{n}')}(v)
R_{(\ud{\ud{1}}\cdots\ud{\ud{n}}''),(\ud{n}'\cdots \ud{1})}^{\mathbf{c}_{(\ud{\ud{1}}\cdots\ud{\ud{n}}'')},-\mathbf{c}_{(\ud{n}'\cdots \ud{1})}}(v)\\
=&\left[R_{(1\cdots n),(\ud{1}\cdots\ud{n}')}^{\mb{c}_{(1\cdots n)},\mb{c}_{(\ud{1}\cdots\ud{n}')}}(u-v)
R_{(1\cdots n),(\ud{\ud{1}}\cdots\ud{\ud{n}}'')}^{\mathbf{c}_{(1\cdots n)},\mathbf{c}_{(\ud{\ud{1}}\cdots\ud{\ud{n}}'')}}(u)
R_{(\ud{1}\cdots \ud{n}'),(\ud{\ud{1}}\cdots\ud{\ud{n}}'')}^{\mathbf{c}_{(\ud{1}\cdots \ud{n}')},\mathbf{c}_{(\ud{\ud{1}}\cdots\ud{\ud{n}}'')}}(v)\right]K^{-,\mathbf{c}_{(1\cdots n)}}_{(1\cdots n)}(u)R_{(\ud{1}\cdots\ud{n}'),(n\cdots 1)}^{\mb{c}_{(\ud{1}\cdots\ud{n}')},-\mb{c}_{(n\cdots 1)}}(u+v) \\
&
K^{-,\mathbf{c}_{(\ud{1}\cdots \ud{n}')}}_{(\ud{1}\cdots \ud{n}')}(v)
R_{(\ud{\ud{1}}\cdots\ud{\ud{n}}''),(n\cdots 1)}^{\mathbf{c}_{(\ud{\ud{1}}\cdots\ud{\ud{n}}'')},-\mathbf{c}_{(n\cdots 1)}}(u)
R_{(\ud{\ud{1}}\cdots\ud{\ud{n}}''),(\ud{n}'\cdots \ud{1})}^{\mathbf{c}_{(\ud{\ud{1}}\cdots\ud{\ud{n}}'')},-\mathbf{c}_{(\ud{n}'\cdots \ud{1})}}(v)\\
=&R_{(\ud{1}\cdots \ud{n}'),(\ud{\ud{1}}\cdots\ud{\ud{n}}'')}^{\mathbf{c}_{(\ud{1}\cdots \ud{n}')},\mathbf{c}_{(\ud{\ud{1}}\cdots\ud{\ud{n}}'')}}(v)
R_{(1\cdots n),(\ud{\ud{1}}\cdots\ud{\ud{n}}'')}^{\mathbf{c}_{(1\cdots n)},\mathbf{c}_{(\ud{\ud{1}}\cdots\ud{\ud{n}}'')}}(u)
\left[R_{(1\cdots n),(\ud{1}\cdots\ud{n}')}^{\mb{c}_{(1\cdots)},\mb{c}_{(\ud{1}\cdots\ud{n}')}}(u-v)K^{-,\mathbf{c}_{(1\cdots n)}}_{(1\cdots n)}(u)R_{(\ud{1}\cdots\ud{n}'),(n\cdots 1)}^{\mb{c}_{(\ud{1}\cdots\ud{n}')},-\mb{c}_{(n\cdots 1)}}(u+v)\right.\\
&\left. 
K^{-,\mathbf{c}_{(\ud{1}\cdots \ud{n}')}}_{(\ud{1}\cdots \ud{n}')}(v)\right]
R_{(\ud{\ud{1}}\cdots\ud{\ud{n}}''),(n\cdots 1)}^{\mathbf{c}_{(\ud{\ud{1}}\cdots\ud{\ud{n}}'')},-\mathbf{c}_{(n\cdots 1)}}(u)
R_{(\ud{\ud{1}}\cdots\ud{\ud{n}}''),(\ud{n}'\cdots \ud{1})}^{\mathbf{c}_{(\ud{\ud{1}}\cdots\ud{\ud{n}}'')},-\mathbf{c}_{(\ud{n}'\cdots \ud{1})}}(v)\\
=&R_{(\ud{1}\cdots \ud{n}'),(\ud{\ud{1}}\cdots\ud{\ud{n}}'')}^{\mathbf{c}_{(\ud{1}\cdots \ud{n}')},\mathbf{c}_{(\ud{\ud{1}}\cdots\ud{\ud{n}}'')}}(v)
K^{-,\mathbf{c}_{(\ud{1}\cdots \ud{n}')}}_{(\ud{1}\cdots \ud{n}')}(v)
R_{(1\cdots n),(\ud{\ud{1}}\cdots\ud{\ud{n}}'')}^{\mathbf{c}_{(1\cdots n)},\mathbf{c}_{(\ud{\ud{1}}\cdots\ud{\ud{n}}'')}}(u)
R_{(1\cdots n),(\ud{n}'\cdots\ud{1})}^{\mb{c}_{(1\cdots n)},-\mb{c}_{(\ud{n}'\cdots\ud{1})}}(u+v)
K^{-,\mathbf{c}_{(1\cdots n)}}_{(1\cdots n)}(u)\\
&\left[R_{(\ud{n}'\cdots\ud{1}),(n\cdots 1)}^{-\mb{c}_{(\ud{n}'\cdots\ud{1})},-\mb{c}_{(n\cdots 1)}}(u-v)
R_{(\ud{\ud{1}}\cdots\ud{\ud{n}}''),(n\cdots 1)}^{\mathbf{c}_{(\ud{\ud{1}}\cdots\ud{\ud{n}}'')},-\mathbf{c}_{(n\cdots 1)}}(u)
R_{(\ud{\ud{1}}\cdots\ud{\ud{n}}''),(\ud{n}'\cdots \ud{1})}^{\mathbf{c}_{(\ud{\ud{1}}\cdots\ud{\ud{n}}'')},-\mathbf{c}_{(\ud{n}'\cdots \ud{1})}}(v)
\right]\\
&=R_{(\ud{1}\cdots \ud{n}'),(\ud{\ud{1}}\cdots\ud{\ud{n}}'')}^{\mathbf{c}_{(\ud{1}\cdots \ud{n}')},\mathbf{c}_{(\ud{\ud{1}}\cdots\ud{\ud{n}}'')}}(v)
K^{-,\mathbf{c}_{(\ud{1}\cdots \ud{n}')}}_{(\ud{1}\cdots \ud{n}')}(v)
\left[R_{(1\cdots n),(\ud{\ud{1}}\cdots\ud{\ud{n}}'')}^{\mathbf{c}_{(1\cdots n)},\mathbf{c}_{(\ud{\ud{1}}\cdots\ud{\ud{n}}'')}}(u)
R_{(1\cdots n),(\ud{n}'\cdots\ud{1})}^{\mb{c}_{(1\cdots n)},-\mb{c}_{(\ud{n}'\cdots\ud{1})}}(u+v)R_{(\ud{\ud{1}}\cdots\ud{\ud{n}}''),(\ud{n}'\cdots \ud{1})}^{\mathbf{c}_{(\ud{\ud{1}}\cdots\ud{\ud{n}}'')},-\mathbf{c}_{(\ud{n}'\cdots \ud{1})}}(v)\right]\\
&K^{-,\mathbf{c}_{(1\cdots n)}}_{(1\cdots n)}(u)R_{(\ud{\ud{1}}\cdots\ud{\ud{n}}''),(n\cdots 1)}^{\mathbf{c}_{(\ud{\ud{1}}\cdots\ud{\ud{n}}'')},-\mathbf{c}_{(n\cdots 1)}}(u)R_{(\ud{n}'\cdots\ud{1}),(n\cdots 1)}^{-\mb{c}_{(\ud{n}'\cdots\ud{1})},-\mb{c}_{(n\cdots 1)}}(u-v)\\
&=U_{(\ud{1}\cdots\ud{n}')}(v)
R_{(1\cdots n),(\ud{n}'\cdots\ud{1})}^{\mb{c}_{(1\cdots n)},-\mb{c}_{(\ud{n}'\cdots\ud{1})}}(u+v)
U_{(1\cdots n)}(u)
R_{(\ud{n}'\cdots\ud{1}),(n\cdots 1)}^{\mb{c}_{(\ud{n}'\cdots\ud{1})},\mb{c}_{(n\cdots 1)}}(u-v),
\end{split}
\end{equation}
where in the derivation we have successively used the fused YBE (\ref{fYBE}) four times and the fused RE (\ref{fre}) once:
YBE$\rightarrow$YBE$\rightarrow$RE$\rightarrow$YBE$\rightarrow$YBE. 

\begin{figure}[h]
 \begin{center}
   \includegraphics[width=1\linewidth]{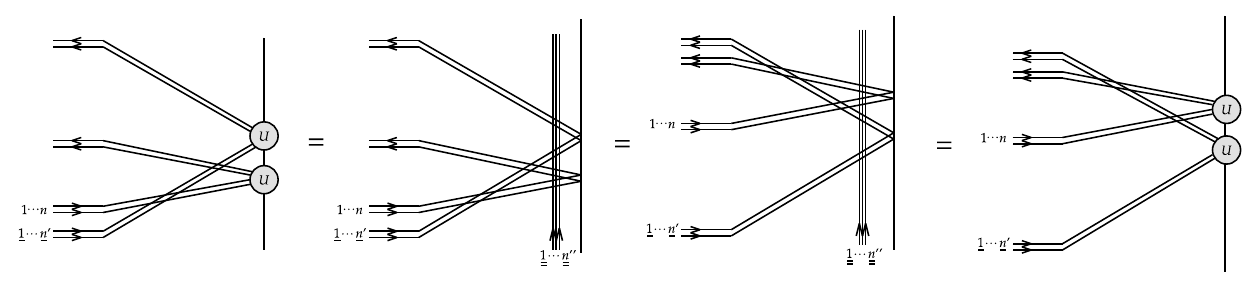}
 \end{center}
\caption{Reflection equation for the fused double-row monodromy matrix $U_{(1\cdots n)}$. \label{fig:fdrmre}}
\end{figure}

The graph representation of the RE for $U_{(1\cdots n)}(u)$ is shown in Fig.~\ref{fig:fdrmre}.

\subsubsection{Fused transfer matrices and the commutativity property}
\begin{figure}[h]
 \begin{center}
   \includegraphics[width=0.9\linewidth]{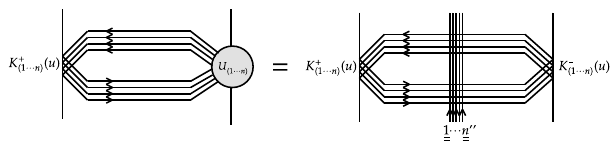}
 \end{center}
\caption{Fused transfer matrix. \label{fig:ftm}}
\end{figure}

Using the fused double-row monodromy matrix $U_{(1\cdots n)}(u)$ (\ref{fdrm}) and the fused $K^+$-matrix introduced above, we can construct the fused transfer matrix as follows:
\begin{equation}\label{ftm}
{\tau}^{\mb{c}_{(1\cdots n)}}(u)\equiv {\tau}^{\mb{c}_{(1\cdots n)},\mb{c}_{(\ud{\ud{1}}\cdots\ud{\ud{n}}'')}}_{(\ud{\ud{1}}\cdots\ud{\ud{n}}'')}(u)=\tr_{(1\cdots n)} K^{+,-\mb{c}_{(1\cdots n)}}_{(1\cdots n)}(u)U_{(1\cdots n),(\ud{\ud{1}}\cdots\ud{\ud{n}}'')}(u),
\end{equation}
where the trace is taken over the auxiliary tensor space $V^{\otimes n}=V_1\otimes\cdots\otimes V_n$ and $\tau^{\mb{c}_{(1\cdots n)}}(u)$ is still an operator in the internal quantum space $V^{\otimes n''}$: ${\tau}^{\mb{c}_{(1\cdots n)},\mb{c}_{(\ud{\ud{1}}\cdots\ud{\ud{n}}'')}}_{(\ud{\ud{1}}\cdots\ud{\ud{n}}'')}(u)\in {\rm{End}}(V_{\ud{\ud{1}}}\otimes\cdots\otimes V_{\ud{\ud{n}}''})$. Graphically, the fused transfer matrix can be represented as a closed loop as shown in Fig.~\ref{fig:ftm}.

Now we will demonstrate the commutativity property of the fused transfer matrices:
\begin{equation}
\left[{\tau}^{\mb{c}_{(1\cdots n)}}(u),{\tau}^{\mb{c}_{(\ud{1}\cdots \ud{n}')}}(v)\right]=0,\quad \forall u,v,c_1\cdots c_n,c_{\ud{1}}\cdots c_{\ud{n}'}\in \mathbb{C}.
\end{equation}
The proof presented below essentially follows the procedure given by Sklyanin in his seminal work on unfused open spin chains \cite{Sklyanin:1988yz}, but is carefully adapted to the fused case. First note the following simple fact which will be used repeatedly throughout the proof:
\begin{equation}\label{eq1}
\tr_V(A^tB^t)=\tr_V(BA)^t=\tr_V(BA)=\tr_V(AB),\quad \forall A,B\in {\rm{End}(V)}.
\end{equation}
Then we begin with the product ${\tau}^{\mb{c}_{(1\cdots n)}}(u){\tau}^{\mb{c}_{(\ud{1}\cdots \ud{n}')}}(v)$ and by using (\ref{ftm}) and (\ref{eq1}) we obtain:
\begin{equation}
\begin{split}
&{\tau}^{\mb{c}_{(1\cdots n)}}(u){\tau}^{\mb{c}_{(\ud{1}\cdots \ud{n}')}}(v)\\
=&\tr_{(1\cdots n)}K^{+,-\mb{c}_{(1\cdots n)}}_{(1\cdots n)}(u)U_{(1\cdots n)}(u)\,\tr_{(\ud{1}\cdots \ud{n}')}K^{+,-\mb{c}_{(\ud{1}\cdots \ud{n}')}}_{(\ud{1}\cdots \ud{n}')}(v)U_{(\ud{1}\cdots \ud{n}')}(v)\\
=&\tr_{(1\cdots n)}K^{+,-\mb{c}_{(1\cdots n)}}_{(1\cdots n)}(u)^{t_{(1\cdots n)}}U_{(1\cdots n)}(u)^{t_{(1\cdots n)}}\,\tr_{(\ud{1}\cdots \ud{n}')}K^{+,-\mb{c}_{(\ud{1}\cdots \ud{n}')}}_{(\ud{1}\cdots \ud{n}')}(v)U_{(\ud{1}\cdots \ud{n}')}(v)\\
=&\tr_{\begin{subarray}{l}(1\cdots n)\\(\ud{1}\cdots\ud{n}')\end{subarray}}K^{+,-\mb{c}_{(1\cdots n)}}_{(1\cdots n)}(u)^{t_{(1\cdots n)}} K^{+,-\mb{c}_{(\ud{1}\cdots \ud{n}')}}_{(\ud{1}\cdots \ud{n}')}(v) U_{(1\cdots n)}(u)^{t_{(1\cdots n)}} U_{(\ud{1}\cdots \ud{n}')}(v)=\cdots
\end{split}
\end{equation}
By inserting the fused crossing unitary relation (\ref{fcuc}) and reorganizing certain terms through transpose operation, we then obtain:

\begin{equation}
\begin{split}
\cdots=&\tr_{(1\cdots n),(\ud{1}\cdots\ud{n}')}K^{+,-\mb{c}_{(1\cdots n)}}_{(1\cdots n)}(u)^{t_{(1\cdots n)}}
K^{+,-\mb{c}_{(\ud{1}\cdots \ud{n}')}}_{(\ud{1}\cdots \ud{n}')}(v)
R_{(n\cdots 1),(\ud{1}\cdots\ud{n}')}^{-\mb{c}_{(n\cdots 1)},\mb{c}_{(\ud{1}\cdots\ud{n}')}}(-u-v+N)^{t_{(1\cdots n)}}\\
&R_{(\ud{1}\cdots\ud{n}'),(n\cdots 1)}^{\mb{c}_{(\ud{1}\cdots\ud{n}')},-\mb{c}_{(n\cdots 1)}}(v+u)^{t_{(1\cdots n)}}
U_{(1\cdots n)}(u)^{t_{(1\cdots n)}} U_{(\ud{1}\cdots \ud{n}')}(v)\\
=&\tr_{\begin{subarray}{l}(1\cdots n)\\(\ud{1}\cdots\ud{n}')\end{subarray}}K^{+,-\mb{c}_{(1\cdots n)}}_{(1\cdots n)}(u)^{t_{(1\cdots n)}}
\left[R_{(n\cdots 1),(\ud{1}\cdots\ud{n}')}^{-\mb{c}_{(n\cdots 1)},\mb{c}_{(\ud{1}\cdots\ud{n}')}}(-u-v+N)
^{\begin{subarray}{l} t_{(\ud{1}\cdots \ud{n}')}\\t_{(1\cdots n)}\end{subarray}}
K^{+,-\mb{c}_{(\ud{1}\cdots \ud{n}')}}_{(\ud{1}\cdots \ud{n}')}(v)^{t_{(\ud{1}\cdots\ud{n}')}}\right]^{t_{(\ud{1}\cdots\ud{n}')}}\\
&\left[U_{(1\cdots n)}(u) R_{(\ud{1}\cdots\ud{n}'),(n\cdots 1)}^{\mb{c}_{(\ud{1}\cdots\ud{n}')},-\mb{c}_{(n\cdots 1)}}(v+u)\right]^{t_{(1\cdots n)}}
U_{(\ud{1}\cdots \ud{n}')}(v)=\cdots
\end{split}
\end{equation}
Owing to the P- and T-symmetry conditions (\ref{psym}) and (\ref{tsym3}), we find:
\begin{equation}\label{eq2}
R_{(n\cdots 1),(\ud{1}\cdots\ud{n}')}^{-\mb{c}_{(n\cdots 1)},\mb{c}_{(\ud{1}\cdots\ud{n}')}}(-u-v+N)
^{t_{(1\cdots n)},t_{(\ud{1}\cdots \ud{n}')}}=R_{(\ud{1}\cdots\ud{n}'),(n\cdots 1)}^{-\mb{c}_{(\ud{1}\cdots\ud{n}')},\mb{c}_{(n\cdots 1)}}(-u-v+N).
\end{equation}
Another relation concerning the transpose operation that we will use is: 
\begin{equation}\label{eq3}
\begin{split}
&\tr_{V_1\otimes V_2}\left(A\otimes B\right)\left(C\otimes D\right)=\tr_{V_1\otimes V_2}\left(AC\otimes BD\right)=\tr_{V_1}(AC)\tr_{V_2}(BD)\\
=&\tr_{V_1}(A^{t_1}C^{t_1})\tr_{V_2}(BD)=\tr_{V_1\otimes V_2}\left(A^{t_1}\otimes B\right)\left(C^{t_1}\otimes D\right)\\
=&\tr_{V_1\otimes V_2}\left(A\otimes B\right)^{t_1}\left(C\otimes D\right)^{t_1}.\qquad \forall A,C\in {\rm{End}}(V_1),\quad \forall B,D\in {\rm{End}}(V_2)
\end{split}
\end{equation}
Using the above equalities (\ref{eq2}) and (\ref{eq3}), the proof proceeds as follows:
\begin{equation}
\begin{split}
\cdots=&\tr_{\begin{subarray}{l}(1\cdots n)\\(\ud{1}\cdots\ud{n}')\end{subarray}}\left[K^{+,-\mb{c}_{(1\cdots n)}}_{(1\cdots n)}(u)^{t_{(1\cdots n)}}
R_{(\ud{1}\cdots\ud{n}'),(n\cdots 1)}^{-\mb{c}_{(\ud{1}\cdots\ud{n}')},\mb{c}_{(n\cdots 1)}}(-u-v+N)
K^{+,-\mb{c}_{(\ud{1}\cdots \ud{n}')}}_{(\ud{1}\cdots \ud{n}')}(v)^{t_{(\ud{1}\cdots\ud{n}')}}\right]^{t_{(\ud{1}\cdots\ud{n}')}}\\
&\left[U_{(1\cdots n)}(u) R_{(\ud{1}\cdots\ud{n}'),(n\cdots 1)}^{\mb{c}_{(\ud{1}\cdots\ud{n}')},-\mb{c}_{(n\cdots 1)}}(v+u)
U_{(\ud{1}\cdots \ud{n}')}(v)\right]^{t_{(1\cdots n)}}\\
=&\tr_{\begin{subarray}{l}(1\cdots n)\\(\ud{1}\cdots\ud{n}')\end{subarray}}\left[K^{+,-\mb{c}_{(1\cdots n)}}_{(1\cdots n)}(u)^{t_{(1\cdots n)}}
R_{(\ud{1}\cdots\ud{n}'),(n\cdots 1)}^{-\mb{c}_{(\ud{1}\cdots\ud{n}')},\mb{c}_{(n\cdots 1)}}(-u-v+N)
K^{+,-\mb{c}_{(\ud{1}\cdots \ud{n}')}}_{(\ud{1}\cdots \ud{n}')}(v)^{t_{(\ud{1}\cdots\ud{n}')}}\right]
^{\begin{subarray}{l}t_{(1\cdots n)}\\t_{(\ud{1}\cdots\ud{n}')}\end{subarray}}\\
&\left[U_{(1\cdots n)}(u) R_{(\ud{1}\cdots\ud{n}'),(n\cdots 1)}^{\mb{c}_{(\ud{1}\cdots\ud{n}')},-\mb{c}_{(n\cdots 1)}}(v+u)
U_{(\ud{1}\cdots \ud{n}')}(v)\right]=\cdots
\end{split}
\end{equation}
Now we insert the unitary condition (\ref{fuc}) into the above equation and use the T-symmetry condition (\ref{tsym3}) together with the transpose operation to obtain:
\begin{equation}
\begin{split}
\cdots=&\rho^{-1}\tr_{\begin{subarray}{l}(1\cdots n)\\(\ud{1}\cdots\ud{n}')\end{subarray}}\left[K^{+,-\mb{c}_{(1\cdots n)}}_{(1\cdots n)}(u)^{t_{(1\cdots n)}}
R_{(\ud{1}\cdots\ud{n}'),(n\cdots 1)}^{-\mb{c}_{(\ud{1}\cdots\ud{n}')},\mb{c}_{(n\cdots 1)}}(-u-v+N)
K^{+,-\mb{c}_{(\ud{1}\cdots \ud{n}')}}_{(\ud{1}\cdots \ud{n}')}(v)^{t_{(\ud{1}\cdots\ud{n}')}}\right]
^{\begin{subarray}{l}t_{(1\cdots n)}\\t_{(\ud{1}\cdots\ud{n}')}\end{subarray}}\\
&R_{(\ud{1}\cdots\ud{n}'),(1\cdots n)}^{\mb{c}_{(\ud{1}\cdots\ud{n}')},\mb{c}_{(1\cdots n)}}(v-u)
R_{(1\cdots n),(\ud{1}\cdots\ud{n}')}^{\mb{c}_{(1\cdots n)},\mb{c}_{(\ud{1}\cdots\ud{n}')}}(u-v)\left[U_{(1\cdots n)}(u) R_{(\ud{1}\cdots\ud{n}'),(n\cdots 1)}^{\mb{c}_{(\ud{1}\cdots\ud{n}')},-\mb{c}_{(n\cdots 1)}}(v+u)
U_{(\ud{1}\cdots \ud{n}')}(v)\right]\\
=&\rho^{-1}\tr_{\begin{subarray}{l}(1\cdots n)\\(\ud{1}\cdots\ud{n}')\end{subarray}}
\left[R_{(\ud{1}\cdots\ud{n}'),(1\cdots n)}^{\mb{c}_{(\ud{1}\cdots\ud{n}')},\mb{c}_{(1\cdots n)}}(v-u)^{\begin{subarray}{l}t_{(1\cdots n)}\\t_{(\ud{1}\cdots\ud{n}')}\end{subarray}}
K^{+,-\mb{c}_{(1\cdots n)}}_{(1\cdots n)}(u)^{t_{(1\cdots n)}}
R_{(\ud{1}\cdots\ud{n}'),(n\cdots 1)}^{-\mb{c}_{(\ud{1}\cdots\ud{n}')},\mb{c}_{(n\cdots 1)}}(-u-v+N)\right.\\
&\left.K^{+,-\mb{c}_{(\ud{1}\cdots \ud{n}')}}_{(\ud{1}\cdots \ud{n}')}(v)^{t_{(\ud{1}\cdots\ud{n}')}}\right.\bigg]^{\begin{subarray}{l}t_{(1\cdots n)}\\t_{(\ud{1}\cdots\ud{n}')}\end{subarray}}R_{(1\cdots n),(\ud{1}\cdots\ud{n}')}^{\mb{c}_{(1\cdots n)},\mb{c}_{(\ud{1}\cdots\ud{n}')}}(u-v)U_{(1\cdots n)}(u) R_{(\ud{1}\cdots\ud{n}'),(n\cdots 1)}^{\mb{c}_{(\ud{1}\cdots\ud{n}')},-\mb{c}_{(n\cdots 1)}}(v+u)
U_{(\ud{1}\cdots \ud{n}')}(v)\\
=&\rho^{-1}\tr_{\begin{subarray}{l}(1\cdots n)\\(\ud{1}\cdots\ud{n}')\end{subarray}}
\left[R_{(1\cdots n),(\ud{1}\cdots\ud{n}')}^{-\mb{c}_{(1\cdots n)},-\mb{c}_{(\ud{1}\cdots\ud{n}')}}(v-u)
K^{+,-\mb{c}_{(1\cdots n)}}_{(1\cdots n)}(u)^{t_{(1\cdots n)}}
R_{(\ud{1}\cdots\ud{n}'),(n\cdots 1)}^{-\mb{c}_{(\ud{1}\cdots\ud{n}')},\mb{c}_{(n\cdots 1)}}(-u-v+N)\right.\\
&\left.K^{+,-\mb{c}_{(\ud{1}\cdots \ud{n}')}}_{(\ud{1}\cdots \ud{n}')}(v)^{t_{(\ud{1}\cdots\ud{n}')}}\right]^{\begin{subarray}{l}t_{(1\cdots n)}\\t_{(\ud{1}\cdots\ud{n}')}\end{subarray}}R_{(1\cdots n),(\ud{1}\cdots\ud{n}')}^{\mb{c}_{(1\cdots n)},\mb{c}_{(\ud{1}\cdots\ud{n}')}}(u-v)U_{(1\cdots n)}(u) R_{(\ud{1}\cdots\ud{n}'),(n\cdots 1)}^{\mb{c}_{(\ud{1}\cdots\ud{n}')},-\mb{c}_{(n\cdots 1)}}(v+u)
U_{(\ud{1}\cdots \ud{n}')}(v)\\
=&\cdots
\end{split}
\end{equation}
where we use $\rho$ to denote $\rho_{(\ud{1}\cdots\ud{n}'),(1\cdots n)}^{\mb{c}_{(\ud{1}\cdots\ud{n}')},\mb{c}_{(1\cdots n)}}(v-u)$ for brevity. Then we use two reflection equations (\ref{fdre}) and (\ref{drre}) for the fused $K^+$-matrices and the fused double-row monodromy matrices to obtain:

\begin{equation}
\begin{split}
\cdots=&\rho^{-1}\tr_{(1\cdots n),(\ud{1}\cdots\ud{n}')}
\left[K^{+,-\mb{c}_{(\ud{1}\cdots \ud{n}')}}_{(\ud{1}\cdots \ud{n}')}(v)^{t_{(\ud{1}\cdots\ud{n}')}}R_{(1\cdots n),(\ud{n}'\cdots\ud{1})}^{-\mb{c}_{(1\cdots n)},\mb{c}_{(\ud{n}'\cdots\ud{1})}}(-u-v+N)
K^{+,-\mb{c}_{(1\cdots n)}}_{(1\cdots n)}(u)^{t_{(1\cdots n)}}
\right.\\
&\left.R_{(\ud{n}'\cdots\ud{1}),(n\cdots 1)}^{\mb{c}_{(\ud{n}'\cdots\ud{1})},\mb{c}_{(n\cdots 1)}}(v-u)\right]^{\begin{subarray}{l}t_{(1\cdots n)}\\t_{(\ud{1}\cdots\ud{n}')}\end{subarray}}
U_{(\ud{1}\cdots \ud{n}')}(v) R_{(1\cdots n),(\ud{n}'\cdots\ud{1})}^{\mb{c}_{(1\cdots n)},-\mb{c}_{(\ud{n}'\cdots\ud{1})}}(u+v)U_{(1\cdots n)}(u) R_{(\ud{n}'\cdots\ud{1}),(n\cdots 1)}^{-\mb{c}_{(\ud{n}'\cdots\ud{1})},-\mb{c}_{(n\cdots 1)}}(u-v)\\
=&\rho^{-1}\tr_{(1\cdots n),(\ud{1}\cdots\ud{n}')}
\left[R_{(\ud{n}'\cdots\ud{1}),(n\cdots 1)}^{\mb{c}_{(\ud{n}'\cdots\ud{1})},\mb{c}_{(n\cdots 1)}}(v-u)\right]^{\begin{subarray}{l}t_{(1\cdots n)}\\t_{(\ud{1}\cdots\ud{n}')}\end{subarray}}
\left[K^{+,-\mb{c}_{(\ud{1}\cdots \ud{n}')}}_{(\ud{1}\cdots \ud{n}')}(v)^{t_{(\ud{1}\cdots\ud{n}')}}R_{(1\cdots n),(\ud{n}'\cdots\ud{1})}^{-\mb{c}_{(1\cdots n)},\mb{c}_{(\ud{n}'\cdots\ud{1})}}(-u-v+N)\right.\\
&\left.K^{+,-\mb{c}_{(1\cdots n)}}_{(1\cdots n)}(u)^{t_{(1\cdots n)}}\right]^{\begin{subarray}{l}t_{(1\cdots n)}\\t_{(\ud{1}\cdots\ud{n}')}\end{subarray}}
U_{(\ud{1}\cdots \ud{n}')}(v) R_{(1\cdots n),(\ud{n}'\cdots\ud{1})}^{\mb{c}_{(1\cdots n)},-\mb{c}_{(\ud{n}'\cdots\ud{1})}}(u+v)U_{(1\cdots n)}(u) R_{(\ud{n}'\cdots\ud{1}),(n\cdots 1)}^{-\mb{c}_{(\ud{n}'\cdots\ud{1})},-\mb{c}_{(n\cdots 1)}}(u-v)\\
=&\cdots
\end{split}
\end{equation}
Proceeding with the P- and T-symmetry conditions (\ref{psym}) and (\ref{tsym3}), the cyclic property of the trace and the unitary condition (\ref{fuc}), we derive:
\begin{equation}
\begin{split}
\cdots=&\tr_{\begin{subarray}{l}(1\cdots n)\\(\ud{1}\cdots\ud{n}')\end{subarray}}
\left[K^{+,-\mb{c}_{(\ud{1}\cdots \ud{n}')}}_{(\ud{1}\cdots \ud{n}')}(v)^{t_{(\ud{1}\cdots\ud{n}')}}R_{(1\cdots n),(\ud{n}'\cdots\ud{1})}^{-\mb{c}_{(1\cdots n)},\mb{c}_{(\ud{n}'\cdots\ud{1})}}(-u-v+N)K^{+,-\mb{c}_{(1\cdots n)}}_{(1\cdots n)}(u)^{t_{(1\cdots n)}}\right]^{\begin{subarray}{l}t_{(1\cdots n)}\\t_{(\ud{1}\cdots\ud{n}')}\end{subarray}}\\
&U_{(\ud{1}\cdots \ud{n}')}(v) R_{(1\cdots n),(\ud{n}'\cdots\ud{1})}^{\mb{c}_{(1\cdots n)},-\mb{c}_{(\ud{n}'\cdots\ud{1})}}(u+v)U_{(1\cdots n)}(u)=\cdots 
\end{split}
\end{equation}
where we have used the relation: $\rho_{(\ud{1}\cdots\ud{n}'),(1\cdots n)}^{\mb{c}_{(\ud{1}\cdots\ud{n}')},\mb{c}_{(1\cdots n)}}(v-u)=
\rho_{(1\cdots n),(\ud{1}\cdots\ud{n}')}^{\mb{c}_{(1\cdots n)},\mb{c}_{(\ud{1}\cdots\ud{n}')}}(u-v)$. 

The last several steps are:
\begin{equation}
\begin{split}
\cdots=&\tr_{\begin{subarray}{l}(1\cdots n)\\(\ud{1}\cdots\ud{n}')\end{subarray}}
\left[K^{+,-\mb{c}_{(\ud{1}\cdots \ud{n}')}}_{(\ud{1}\cdots \ud{n}')}(v)^{t_{(\ud{1}\cdots\ud{n}')}}R_{(1\cdots n),(\ud{n}'\cdots\ud{1})}^{-\mb{c}_{(1\cdots n)},\mb{c}_{(\ud{n}'\cdots\ud{1})}}(-u-v+N)K^{+,-\mb{c}_{(1\cdots n)}}_{(1\cdots n)}(u)^{t_{(1\cdots n)}}\right]^{t_{(1\cdots n)}}\\
&\left[U_{(\ud{1}\cdots \ud{n}')}(v) R_{(1\cdots n),(\ud{n}'\cdots\ud{1})}^{\mb{c}_{(1\cdots n)},-\mb{c}_{(\ud{n}'\cdots\ud{1})}}(u+v)U_{(1\cdots n)}(u)\right]^{t_{(\ud{1}\cdots\ud{n}')}}\\
=&\tr_{\begin{subarray}{l}(1\cdots n)\\(\ud{1}\cdots\ud{n}')\end{subarray}} K^{+,-\mb{c}_{(\ud{1}\cdots \ud{n}')}}_{(\ud{1}\cdots \ud{n}')}(v)^{t_{(\ud{1}\cdots\ud{n}')}} K^{+,-\mb{c}_{(1\cdots n)}}_{(1\cdots n)}(u)\left[R_{(1\cdots n),(\ud{n}'\cdots\ud{1})}^{-\mb{c}_{(1\cdots n)},\mb{c}_{(\ud{n}'\cdots\ud{1})}}(-u-v+N)\right]^{t_{(1\cdots n)}}\\
&\left[R_{(1\cdots n),(\ud{n}'\cdots\ud{1})}^{\mb{c}_{(1\cdots n)},-\mb{c}_{(\ud{n}'\cdots\ud{1})}}(u+v)\right]^{t_{(\ud{1}\cdots\ud{n}')}}
U_{(\ud{1}\cdots \ud{n}')}(v)^{t_{(\ud{1}\cdots\ud{n}')}}U_{(1\cdots n)}(u)\\
=&\tr_{(1\cdots n),(\ud{1}\cdots\ud{n}')}
K^{+,-\mb{c}_{(\ud{1}\cdots \ud{n}')}}_{(\ud{1}\cdots \ud{n}')}(v)^{t_{(\ud{1}\cdots\ud{n}')}}
U_{(\ud{1}\cdots \ud{n}')}(v)^{t_{(\ud{1}\cdots\ud{n}')}}
K^{+,-\mb{c}_{(1\cdots n)}}_{(1\cdots n)}(u)U_{(1\cdots n)}(u)\\
=&{\tau}^{\mb{c}_{(\ud{1}\cdots \ud{n}')}}(v){\tau}^{\mb{c}_{(1\cdots n)}}(u),
\end{split}
\end{equation}
where the fused crossing unitary condition (\ref{fcuc}) has been employed.  Therefore, we complete the proof of the commutativity property of the fused transfer matrices. Note that throughout the proof, we need to pay much attention to the order and the flipping of the signs of the parameters $\{c_i\}$ and $\{c_{\ud{i}}\}$.

\begin{figure}[h]
 \begin{center}
   \includegraphics[width=0.8\linewidth]{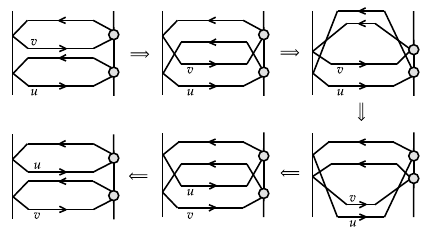}
 \end{center}
\caption{Graphical representation of the commuting relation: $\tau(u)\tau(v)=\tau(v)\tau(u)$. We have successively employed: crossing unitary $\rightarrow$ unitary $\rightarrow$ RE $\rightarrow$ unitary $\rightarrow$ crossing unitary relations.\label{fig:fmc}}
\end{figure}

There is also a nice graph interpretation for the commuting relation of the fused transfer matrices (see, e.g., \cite{Zhou:1995tp} and \cite{Zhou:1995zy}), which will is elaborated again here and illustrated in Fig.~\ref{fig:fmc}. 

\subsection{Evaluations of the fused transfer matrices}
\subsubsection{Fused transfer matrices evaluated in total space}
\begin{figure}[h]
 \begin{center}
   \includegraphics[width=0.85\linewidth]{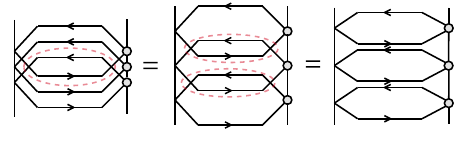}
 \end{center}
\caption{Graph representation of the relation $\tau^{c_1c_2c_3}(u)=\tau^{c_3}(u)\tau^{c_2}(u)\tau^{c_1}(u)$, where the dashed circles indicate the application of crossing unitary conditions. \label{fig:tit}}
\end{figure}

We begin with the evaluation of the fused transfer matrix in the total space $V^{\otimes n}$. By transposing the decomposition relation (\ref{dKp2}) and using the T-symmetry relation (\ref{tsym3}), we obtain:
\begin{equation}
K^{+,-\mb{c}_{(1\cdots n)}}_{(1\cdots n)}(u)=
K^{+,-\mb{c}_{(1\cdots n-1)}}_{(1\cdots n-1)}(u) R_{(n-1\cdots 1),n}^{-\mb{c}_{n-1\cdots 1},c_n}(-2u+N) K^{+,-c_n}_n(u).
\end{equation}
Hence, for the case $n=2$ first, we have:
\begin{equation}\label{eq4}
\begin{split}
\tau^{c_1c_2}(u)=&\tr_{(12)}K^{+,-c_1,-c_2}_{(12)}(u)U_{(12)}(u)\\
=&\tr_{(12)}\left[K^{+,-c_1}_1(u)R_{21}^{-c_2,-c_1}(-2u+N)K^{+,-c_2}_2(u)\right]\left[U_2(u)R_{12}^{c_1,-c_2}(2u)U_1(u)\right]\\
=&\tr_{(12)}K^{+,-c_1}_1(u)K^{+,-c_2}_2(u)^{t_2}R_{21}^{-c_2,-c_1}(-2u+N)^{t_2}R_{12}^{c_1,-c_2}(2u)^{t_2}U_2(u)^{t_2}U_1(u)\\
=&\tr_{(12)}K^{+,-c_1}_1(u)K^{+,-c_2}_2(u)^{t_2}U_2(u)^{t_2}U_1(u)\\
=&\tr_2\left[K^{+,-c_2}_2(u)^{t_2}U_2(u)^{t_2}\right]\tr_1\left[K^{+,-c_1}_1(u)U_1(u)\right]=\tau^{c_2}(u)\tau^{c_1}(u).
\end{split}
\end{equation}
Then we find the $n$-fused transfer matrix $\tau^{\mb{c}_{(1\cdots n)}}(u)$ is simply the multiplication of $n$ fundamental transfer matrices:
\begin{equation}\label{tit}
\tau^{\mb{c}_{(1\cdots n)}}(u)=\tau^{c_n}(u)\tau^{c_{n-1}}(u)\cdots\tau^{c_1}(u),
\end{equation}
which can be proved easily by a similar procedure as in (\ref{eq4}):
\begin{equation}
\begin{split}
\tau^{\mb{c}_{(1\cdots n)}}=&\tr_{(1\cdots n)}\left[K^{+,-\mb{c}_{(1\cdots n-1)}}_{(1\cdots n-1)} R_{(n-1\cdots 1),n}^{-\mb{c}_{(n-1\cdots 1)},c_n}(-2u+N) K^{+,-c_n}_n(u)\right]\\
&\left[U_n(u)R_{(1\cdots n-1),n}^{\mb{c}_{(1\cdots n-1)},-c_n}(2u) U_{(1\cdots n-1)}(u)\right]\\
=&\tr_{(1\cdots n)}K^{+,-\mb{c}_{(1\cdots n-1)}}_{(1\cdots n-1)}K^{+,-c_n}_n(u)^{t_n}R_{(n-1\cdots 1),n}^{-\mb{c}_{(n-1\cdots 1)},c_n}(-2u+N)^{t_n}
R_{(1\cdots n-1),n}^{\mb{c}_{(1\cdots n-1)},-c_n}(2u)^{t_n}\\
&U_n(u)^{t_n}U_{(1\cdots n-1)}(u)\\
=&\left[\tr_{n}K^{+,-c_n}_n(u)^{t_n}U_n(u)^{t_n}\right]\left[\tr_{(1\cdots n-1)}K^{+,-\mb{c}_{(1\cdots n-1)}}_{(1\cdots n-1)}U_{(1\cdots n-1)}(u)\right]
=\tau^{c_n}\tau^{\mb{c}_{(1\cdots n-1)}}.
\end{split}
\end{equation}
The relation (\ref{tit}) can also be understood graphically by successive applications of crossing unitary conditions, as shown in Fig.~\ref{fig:tit} for $n=3$.

\subsubsection{Fused transfer matrices evaluated in the invariant subspace}
By combining the definition (\ref{fdrm}) of $U_{(1\cdots n)}(u)$ with the relations (\ref{Bpm}), (\ref{eq5}), and (\ref{eq6}), we find that the invariant subspace of $U_{(1\cdots n)}(u)$ is given by $W = B^{\mathbf{c}_{(1\cdots n)}}_{(1\cdots n)} V^{\otimes n}$, which coincides precisely with the invariant subspace of $K^{+, -\mathbf{c}_{(1\cdots n)}}_{(1\cdots n)}(u)$. Thus we have:
\begin{equation}
K^{+,-\mb{c}_{(1\cdots n)}}_{(1\cdots n)}(u) U_{(1\cdots n)}\in {\rm{End}}(W).
\end{equation}
Then the trace for the fused transfer matrix can be restricted to the invariant subspace $W$:
\begin{equation}
\tau^W(u)=\tr_W\left(K^{+,-\mb{c}_{(1\cdots n)}}_{(1\cdots n)}(u) U_{(1\cdots n)}(u)\right).
\end{equation}
When $B^{\mathbf{c}_{(1\cdots n)}}_{(1\cdots n)}$ is a projection operator: $V^{\otimes n}={\rm{Im}}B\oplus {\rm{Ker}}B$ (where $B$ denotes $B^{\mathbf{c}_{(1\cdots n)}}_{(1\cdots n)}$ for short), we choose the corresponding basis for $W(={\rm{Im}}B)$ and $W^\bot(={\rm{Ker}}B)$ as:
\begin{equation}
\begin{split}
W&={\rm{{span}}}\{|i\rangle,i=1\cdots{\rm{dim}}W\},\\
W^\bot&={\rm{{span}}}\{|\tilde{\alpha}\rangle,\alpha=1\cdots{\rm{dim}}W^\bot\}.
\end{split}
\end{equation}
Since $B|i\rangle=|i\rangle,B|\tilde{\alpha}\rangle=0$, we obtain:
\begin{equation}\label{rtau}
\begin{split}
\tau^W(u)=&\sum_{i}\langle i|K^{+,-\mb{c}_{(1\cdots n)}}_{(1\cdots n)}(u) U_{(1\cdots n)}(u)|i\rangle
=\sum_{i}\langle i|K^{+,-\mb{c}_{(1\cdots n)}}_{(1\cdots n)}(u) U_{(1\cdots n)}(u)B^{\mathbf{c}_{(1\cdots n)}}_{(1\cdots n)}|i\rangle\\
=&\sum_{i}\langle i|B^{\mathbf{c}_{(1\cdots n)}}_{(1\cdots n)}K^{+,-\mb{c}_{(1\cdots n)}}_{(1\cdots n)}(u)B^{\mathbf{c}_{(1\cdots n)}}_{(1\cdots n)} B^{\mathbf{c}_{(1\cdots n)}}_{(1\cdots n)}U_{(1\cdots n)}(u)B^{\mathbf{c}_{(1\cdots n)}}_{(1\cdots n)}|i\rangle\\
=&\tr_W\left[B^{\mathbf{c}_{(1\cdots n)}}_{(1\cdots n)}K^{+,-\mb{c}_{(1\cdots n)}}_{(1\cdots n)}(u)B^{\mathbf{c}_{(1\cdots n)}}_{(1\cdots n)} B^{\mathbf{c}_{(1\cdots n)}}_{(1\cdots n)}U_{(1\cdots n)}(u)B^{\mathbf{c}_{(1\cdots n)}}_{(1\cdots n)}\right]\\
=&\tr_{W\oplus W^{\bot}}\left[B^{\mathbf{c}_{(1\cdots n)}}_{(1\cdots n)}K^{+,-\mb{c}_{(1\cdots n)}}_{(1\cdots n)}(u)B^{\mathbf{c}_{(1\cdots n)}}_{(1\cdots n)} B^{\mathbf{c}_{(1\cdots n)}}_{(1\cdots n)}U_{(1\cdots n)}(u)B^{\mathbf{c}_{(1\cdots n)}}_{(1\cdots n)}\right],
\end{split}
\end{equation}
where in the derivations, we have used the relations:
\begin{equation}
\begin{split}
&U_{(1\cdots n)}(u)B^{\mathbf{c}_{(1\cdots n)}}_{(1\cdots n)}=B^{\mathbf{c}_{(1\cdots n)}}_{(1\cdots n)}U_{(1\cdots n)}(u)B^{\mathbf{c}_{(1\cdots n)}}_{(1\cdots n)},
\\&K^{+,-\mb{c}_{(1\cdots n)}}_{(1\cdots n)}(u)B^{\mathbf{c}_{(1\cdots n)}}_{(1\cdots n)}=B^{\mathbf{c}_{(1\cdots n)}}_{(1\cdots n)}K^{+,-\mb{c}_{(1\cdots n)}}_{(1\cdots n)}(u)B^{\mathbf{c}_{(1\cdots n)}}_{(1\cdots n)}. 
\end{split}
\end{equation}
For an appropriately selected set of  parameters $\{c_1\cdots c_n\}$, the fusion operator $B^{\mathbf{c}_{(1\cdots n)}}_{(1\cdots n)}$ could be a 1-dimensional projector in $V^{\otimes n}$. For any 1-dim projector $Q$, we have:
\begin{equation}
QTQ=\tr\left(QT\right)Q,\quad \forall T.
\end{equation} 
Hence, the expression of $\tau^{W}(u)$ in (\ref{rtau}) can be further reduced to:
\begin{equation}
\tau^W(u)=\tr_W\left[B^{\mathbf{c}_{(1\cdots n)}}_{(1\cdots n)}K^{+,-\mb{c}_{(1\cdots n)}}_{(1\cdots n)}(u)\right]
\tr_W\left[B^{\mathbf{c}_{(1\cdots n)}}_{(1\cdots n)}U_{(1\cdots n)}(u)\right],
\end{equation}
which equals to the product of two quantum determinants.

\begin{example}
As an illustration, we consider the fusion of two fundamental open spin chain transfer matrices defined on a common quantum space $V_{\ud{1}}\otimes V_{\ud{2}}\otimes V_{\ud{3}}$. The first building block is the double-row monodromy matrix:
\begin{equation}
U_{(12)}(u)=R_{(12),(\ud{1}\ud{2}\ud{3})}^{c_1,c_2,\mb{c}_{(\ud{1}\ud{2}\ud{3})}}(u)K^{-,c_1,c_2}_{(12)}(u)R_{(\ud{1}\ud{2}\ud{3}),(21)}^{\mb{c}_{(\ud{1}\ud{2}\ud{3})},c_2,c_1}(u),
\end{equation}
where the 2-fused $K^-$-matrix is given by (\ref{exf1}) in \ref{ex3}. Note that the fusion only occurs in the auxiliary tensor product space $V_1\otimes V_2$. Then the 2-fused $K^+$-matrix can be obtained from (\ref{kp1}) by an additional transposition:
\begin{equation}
K^{+,-c_1,-c_2}_{(12)}(u)=K_1^{+,-c_1}(u)R_{12}^{-c_1,c_2}(-2u+N)K^{+,-c_2}_2(u).
\end{equation}
Thus, the fused transfer matrix is:
\begin{equation}\label{exf3}
  \tau^{c_1c_2}(u)=\tr_{V_1\otimes V_2}\left(K^{+,-c_1,-c_2}_{(12)}(u)U_{(12)}(u)\right).
\end{equation}
In the symmetric and antisymmetric invariant subspace (defined in \ref{ex3}), the restricted fused transfer matrices are given by:
\begin{equation}\label{exf4}
\begin{split}
\tau^W&=\tr_{V_1\otimes V_2}\left(K^{+,0,-1}_{(12)}(u)U_{(12)}(u)B_{12}^{0,1}\right),\\
\tau^{W^{\perp}}&=\tr_{V_1\otimes V_2}\left(K^{+,0,1}_{(12)}(u)U_{(12)}(u)B_{12}^{0,-1}\right).
\end{split}
\end{equation}
With explicit solutions $K^{\pm}(u)$ of the fundamental reflection equations (\ref{re}) and (\ref{dre}) at hand, the fused transfer matrices in (\ref{exf3}) and (\ref{exf4}) can be calculated directly.

\end{example}
\section{ABJM Alternating Spin Chain: A $\mathfrak{su}(4)$ Fusion Model}
In this section, we consider a specific fusion scheme for $\mathfrak{su}(4)$ spin chains, namely constructing the $R$-matrices acting on the fundamental representation space ``$V_{\mathbf{4}}$" and the anti-fundamental representation (conjugate representation) space  ``$V_{\bar{\mathbf{4}}}$" of $\mathfrak{su}(4)$ \footnote{Here we give some notational conventions: We denote the fundamental representation space $\mathbf{4}$ of $\mathfrak{su}(4)$ by $V_i$ or $V_{\mathbf{4}}$. The reflection matrix defined on this space is written as $K^{-(+)}_i(u)$ or $K^{-(+)}_{\mathbf{4}}(u)$. Similarly, the anti-fundamental representation space $\bar{\mathbf{4}}$ is denoted by $V_{\bar{i}}$ or $V_{\bar{\mathbf{4}}}$, with its reflection matrix expressed as $K^{-(+)}_{\bar{i}}(u)$ or $K^{-(+)}_{\bar{\mathbf{4}}}(u)$. Likewise, the $R$-matrix on $\mathbf{4}\otimes \bar{\mathbf{4}}$ is denoted as $R_{i\bar{j}}(u)$ or $R_{\mathbf{4}\bar{\mathbf{4}}}(u)$.}: 
\begin{equation}\label{eq7}
R_{a\bar{b}}(u)\in {\rm{End}}(V_\mathbf{4}\otimes V_{\bar{\mathbf{4}}}),\quad R_{\bar{a}b}(u)\in {\rm{End}}(V_{\bar{\mathbf{4}}}\otimes V_\mathbf{4}),
\end{equation}
as well as the $K^-$-matrices: 
\begin{equation}\label{eq8}
K^-_{a}(u)\in {\rm{End}}(V_\mathbf{4}),\quad K^-_{\bar{a}}(u)\in {\rm{End}}(V_{\bar{\mathbf{4}}}).
\end{equation}
The anti-fundamental representation $\bar{\mathbf{4}}$ is embedded in the tensor product decomposition:
\begin{equation}
\mathbf{4} \otimes \mathbf{4} \otimes \mathbf{4} = 
\bar{\mathbf{4}} \oplus 
\mathbf{20_S} \oplus 
\mathbf{20}^{(1)}_{\mathbf{M}} \oplus 
\mathbf{20}^{(2)}_{\mathbf{M}},
\end{equation}
where the anti-fundamental representation $\bar{\mathbf{4}}$  corresponds to the fully antisymmetric representation while the other three are the fully symmetric representation $\mathbf{20_S}$ and the two mixed-symmetric representations $\mathbf{20}^{(1)}_{\mathbf{M}}$ and $\mathbf{20}^{(2)}_{\mathbf{M}}$.  The corresponding Young tableau decomposition is:
\begin{equation}
\yng(1) \otimes \yng(1) \otimes \yng(1)=\underbrace{\yng(1,1,1)}_{\bar{\mathbf{4}}} \oplus \,\,\,\yng(3) \oplus \yng(2,1) \oplus\, \yng(2,1).
\end{equation}
According to fusion theorem given in Sec.\ref{Sec:fusion thm}, we only need to set the parameters $c_1,c_2,c_3$ in the fusion operator $B_{(123)}^{c_1c_2c_3}$ as the  contents corresponding to the anti-fundamental representation's Young tableau (denoted by $[1]^3$), i.e., $c_i={\rm{cc}}([1]^3|i),\,i=1,2,3$, to obtain the projection operator from the tensor product space $V_1\otimes V_2\otimes V_3$ of the three fundamental representations onto the anti-fundamental representation space $V_{\bar{1}}$. Concretely, we find $c_1=0,c_2=-1,c_3=-2$ and then $B_{(123)}^{(0,-1,-2)}$ turns out to be:
\begin{equation}
\begin{split}
B_{(123)}^{(0,-1,-2)}=&R_{12}(1)R_{13}(2)R_{23}(1)=\left(\mathbb{I}-\mathbb{P}_{12}\right)\left(\mathbb{I}-\frac{1}{2}\mathbb{P}_{13}\right)\left(\mathbb{I}-\mathbb{P}_{23}\right)\\
=&\mathbb{I}-\mathbb{P}_{12}-\mathbb{P}_{13}-\mathbb{P}_{23}+\mathbb{P}_{12}\mathbb{P}_{13}+\mathbb{P}_{12}\mathbb{P}_{23},
\end{split}
\end{equation}
which is indeed the Young symmetrizer (up to a numerical factor 1/6) to the anti-fundamental representation space $V_{\bar{1}}$:
$V_{\bar{\mathbf{4}}}=B_{(123)}^{(0,-1,-2)}V^{\otimes 3}_{\mathbf{4}}$.

\begin{figure}[h]
 \begin{center}
   \includegraphics[width=0.75\linewidth]{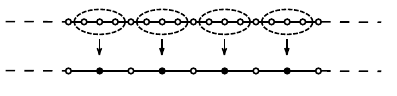}
 \end{center}
\caption{$\mathfrak{su}(4)$ alternating spin chain as a fusion model. \label{fig:alt}}
\end{figure}

Now, starting from a spin chain composed of $\mathfrak{su}(4)$ fundamental representations $\mathbf{4}$, we successively fuse the last three sites within every four adjacent sites into the anti-fundamental representation space $\bar{\mathbf{4}}$. This procedure yields an alternating spin chain with a $\mathbf{4}-\bar{\mathbf{4}}$ staggered pattern, as illustrated in the following diagram Fig.~\ref{fig:alt}.

Next we shall demonstrate that the $R$-matrices (\ref{eq7}) and the $K$-matrices (\ref{eq8}) constructed via the fusion scheme described above are equivalent to those in ABJM theory \cite{Minahan:2008hf,Bak:2008cp}, indicating that the ABJM spin chain is a fusion model.
\subsection{Fused $R$-matrix solutions}
\subsubsection{$R_{\mathbf{4\bar{4}}}(u)$ type solutions}
We first consider solutions of the type $R_{\mathbf{4\bar{4}}}(u)$, which can be obtained by restricting the fused $R$-matrix $R_{1,({\ud{1}},{\ud{2}},{\ud{3}})}(u)\in {\rm{End}}(V_\mathbf{4}^{\otimes4})$ to the invariant subspace $V_\mathbf{4}\otimes V_{\bar{\mathbf{4}}}$:
\begin{equation}
R_{1\bar{1}}(u)=R_{1,(\ud{1},\ud{2},\ud{3})}^{c_1,c_{\ud{1}},c_{\ud{2}},c_{\ud{3}}}(u)\bigg|_{V_{\mathbf{4}}\otimes V_{\bar{\mathbf{4}}}},
\end{equation}
where the parameters are chosen to be $c_1=0,c_{\ud{1}}=0,c_{\ud{2}}=-1,c_{\ud{3}}=-2$ accordingly. We will use the canonical basis for $V_{\mathbf{4}}$: $V_{\mathbf{4}}={\rm{span}}\{e_1,e_2,e_3,e_4\} $. The basis for the subspace $V_{\bar{\mathbf{4}}}$ in $V_\mathbf{4}\otimes V_\mathbf{4}\otimes V_\mathbf{4}$ is chosen to be:
\begin{equation}\label{4barbasis}
\begin{split}
\bar{e}_1=B_{(\ud{1},\ud{2},\ud{3})}^{(0,-1,-2)}e_2\otimes e_3\otimes e_4,\qquad \bar{e}_2=B_{(\ud{1},\ud{2},\ud{3})}^{(0,-1,-2)}e_1\otimes e_4\otimes e_3,\\
\bar{e}_3=B_{(\ud{1},\ud{2},\ud{3})}^{(0,-1,-2)}e_1\otimes e_2\otimes e_4,\qquad \bar{e}_4=B_{(\ud{1},\ud{2},\ud{3})}^{(0,-1,-2)}e_1\otimes e_3\otimes e_2,
\end{split}
\end{equation}
or written as $\bar{e}_i=\varepsilon_{ijkl}B_{(\ud{1},\ud{2},\ud{3})}^{(0,-1,-2)} e_j\otimes e_k\otimes e_l,\,i,j,k,l=1,2,3,4$. Then the basis for the tensor space $V_\mathbf{4}\otimes V_{\bar{\mathbf{4}}}$ is given by:
\begin{equation}
E_{ij}=e_i\otimes \bar{e}_j,\quad i,j=1,2,3,4.
\end{equation}
Due to the intertwining relation
\begin{equation}
R_{1,(\ud{1},\ud{2},\ud{3})}^{0,(0,-1,-2)}(u)B_{(\ud{1},\ud{2},\ud{3})}^{(0,-1,-2)}=B_{(\ud{1},\ud{2},\ud{3})}^{(0,-1,-2)} 
R_{1\ud{3}}(u+2)R_{1\ud{2}}(u+1)R_{1\ud{1}}(u),
\end{equation}
we obtain:
\begin{equation}
\begin{split}
R_{1,(\ud{1},\ud{2},\ud{3})}^{0,(0,-1,-2)}(u) E_{11}&=E_{11}+\frac{1}{u}E_{22}+\frac{1}{u}E_{33}+\frac{1}{u}E_{44},\\
R_{1,(\ud{1},\ud{2},\ud{3})}^{0,(0,-1,-2)}(u) E_{22}&=\frac{1}{u}E_{11}+E_{22}+\frac{1}{u}E_{33}+\frac{1}{u}E_{44},\\
R_{1,(\ud{1},\ud{2},\ud{3})}^{0,(0,-1,-2)}(u) E_{33}&=\frac{1}{u}E_{11}+\frac{1}{u}E_{22}+E_{33}+\frac{1}{u}E_{44},\\
R_{1,(\ud{1},\ud{2},\ud{3})}^{0,(0,-1,-2)}(u) E_{33}&=\frac{1}{u}E_{11}+\frac{1}{u}E_{22}+\frac{1}{u}E_{33}+E_{44},\\
\end{split}
\end{equation}
and $R_{1,(\ud{1},\ud{2},\ud{3})}^{0,(0,-1,-2)}(u) E_{ij}=\frac{u-1}{u}E_{ij}, \,i\neq j$. 
Therefore, up to an overall scalar factor, the fused $R$-matrix restricted to the invariant subspace has the following form:
\begin{equation}\label{eq:fusedR}
R_{a\bar{b}}(u) = (u - 1)\mathbb{I} + \mathbb{K}_{a\bar{b}},
\end{equation}
where $\mathbb{K}_{a\bar{b}} \in \mathrm{End}(V_{\mathbf{4}} \otimes V_{\bar{\mathbf{4}}})$ denotes the \textit{trace operator} defined through the canonical basis:
\begin{equation}
\mathbb{K} =  e_{ij} \otimes \bar{e}_{ij},
\end{equation}
with $\{e_{ij}\} \subset \mathrm{End}(V_{\mathbf{4}})$ and $\{\bar{e}_{ij}\} \subset \mathrm{End}(V_{\bar{\mathbf{4}}})$ being the elementary operator bases satisfying:
\begin{equation}
e_{ij} e_k = \delta_{jk} e_i, \quad \bar{e}_{ij} \bar{e}_k = \delta_{jk} \bar{e}_i .
\end{equation}
\subsubsection{$R_{\bar{\mathbf{4}}\mathbf{4}}(u)$ type solutions}
Analogously, for solutions of the type $R_{\bar{\mathbf{4}}\mathbf{4}}(u)$, they originate from the following restriction:
\begin{equation}
R_{\bar{1}\underline{1}}(u) = R_{(123),\underline{1}}^{(0,-1,-2),0}(u) \bigg|_{V_{\bar{\mathbf{4}}} \otimes V_{\mathbf{4}}}.
\end{equation}
By choosing the basis on $V_{\mathbf{4}}\otimes V_{\bar{\mathbf{4}}}$ as:
\begin{equation}
\tilde{E}_{ij} = \bar{e}_i \otimes e_j, \quad i,j=1,2,3,4,
\end{equation}
and invoking the intertwining relation:
\begin{equation}
R_{(123),\underline{1}}^{(0,-1,-2),0}(u) B_{(123)}^{(0,-1,-2)}=B_{(123)}^{(0,-1,-2)} R_{1\underline{1}}(u)
R_{2\underline{1}}(u-1)R_{3\underline{1}}(u-2),
\end{equation}
we derive the explicit actions:
\begin{equation}
\begin{split}
R_{(123),\underline{1}}^{(0,-1,-2),0}(u)\tilde{E}_{11}=\tilde{E}_{11}+\frac{1}{u-2}\tilde{E}_{22}+\frac{1}{u-2}\tilde{E}_{33}+\frac{1}{u-2}\tilde{E}_{44},\\
R_{(123),\underline{1}}^{(0,-1,-2),0}(u)\tilde{E}_{22}=\frac{1}{u-2}\tilde{E}_{11}+\tilde{E}_{22}+\frac{1}{u-2}\tilde{E}_{33}+\frac{1}{u-2}\tilde{E}_{44},\\
R_{(123),\underline{1}}^{(0,-1,-2),0}(u)\tilde{E}_{33}=\frac{1}{u-2}\tilde{E}_{11}+\frac{1}{u-2}\tilde{E}_{22}+\tilde{E}_{33}+\frac{1}{u-2}\tilde{E}_{44},\\
R_{(123),\underline{1}}^{(0,-1,-2),0}(u)\tilde{E}_{44}=\frac{1}{u-2}\tilde{E}_{11}+\frac{1}{u-2}\tilde{E}_{22}+\frac{1}{u-2}\tilde{E}_{33}+\tilde{E}_{44},
\end{split}
\end{equation}
along with the actions on the off-diagonal basis elements:
\begin{equation}
R_{(123),\underline{1}}^{(0,-1,-2),0}(u)\tilde{E}_{ij}=\frac{u-3}{u-2}\tilde{E}_{ij},\,i\neq j.
\end{equation}
Consequently, up to an overall scalar factor, the solution of type $R_ {\bar{\mathbf{4}}\mathbf{4}}(u)$  is expressed as:
\begin{equation}\label{eqn5}
R_{\bar{a}b}(u) = (u-3)\mathbb{I} + \mathbb{K}_{\bar{a}b}.
\end{equation}
\subsubsection{Equivalence of the fused and ABJM $R$-matrices}
\begin{figure}[h]
 \begin{center}
   \includegraphics[width=0.7\linewidth]{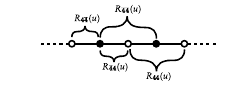}
 \end{center}
\caption{Four types of $R$-matrices in the $\mathfrak{su}(4)$ alternating spin chain. \label{fig:altR}}
\end{figure}

To describe the $\mathfrak{su}(4)$ alternating spin chain, we need four types of $R$-matrices shown in Fig.~\ref{fig:altR}.

One set of the solutions obtained from the above fusion method are:
\begin{eqnarray}\label{fusedRsets}
\left\{
\begin{array}{l}
R_{ab}(u)=u\mathbb{I}_{ab}-\mathbb{P}_{ab},\\
R_{\bar{a}\bar{b}}(u)=u\mathbb{I}_{\bar{a}\bar{b}}-\mathbb{P}_{\bar{a}\bar{b}},\\
R_{a\bar{b}}(u)=(u - 1)\mathbb{I}_{a\bar{b}} + \mathbb{K}_{a\bar{b}},\\
R_{\bar{a}b}(u) = (u-3)\mathbb{I}_{\bar{a}b} + \mathbb{K}_{\bar{a}b}.
\end{array}
\right.
\end{eqnarray}

Another set of $R$-matrix solutions originate from the ABJM spin chain model \cite{Minahan:2008hf,Bak:2008cp}:
\begin{eqnarray}\label{abjmR}
\left\{
\begin{array}{l}
\tilde{R}_{ab}(u)=u\mathbb{I}_{ab}+\mathbb{P}_{ab},\\
\tilde{R}_{\bar{a}\bar{b}}(u)=u\mathbb{I}_{\bar{a}\bar{b}}+\mathbb{P}_{\bar{a}\bar{b}},\\
\tilde{R}_{a\bar{b}}(u)=-(u + 2)\mathbb{I}_{a\bar{b}} + \mathbb{K}_{a\bar{b}},\\
\tilde{R}_{\bar{a}b}(u) = -(u+2)\mathbb{I}_{\bar{a}b} + \mathbb{K}_{\bar{a}b}.
\end{array}
\right.
\end{eqnarray}
The equivalence of these two sets of solutions can be easily demonstrated. As a concrete illustration, we start from the YBE satisfied by the ABJM solutions (\ref{abjmR}), 
\begin{equation}
\tilde{R}_{a\bar{b}}(u-v)\tilde{R}_{ac}(u)\tilde{R}_{\bar{b}c}(v)=\tilde{R}_{\bar{b}c}(v)\tilde{R}_{ac}(u)\tilde{R}_{a\bar{b}}(u-v)\\
\end{equation}
which yileds:
\begin{equation}
\left(-u+v-2+\mathbb{K}_{a\bar{b}}\right)\left(u+\mathbb{P}_{ac}\right)\left(-v-2+\mathbb{K}_{\bar{b}c}\right)
=\left(-v-2+\mathbb{K}_{\bar{b}c}\right)\left(u+\mathbb{P}_{ac}\right)\left(-u+v-2+\mathbb{K}_{a\bar{b}}\right).
\end{equation}
Then letting $u'=-u,\,v'=-v+1$, we obtain:
\begin{equation}
\left(u'-v'-1+\mathbb{K}_{a\bar{b}}\right)\left(u'-\mathbb{P}_{ac}\right)\left(v'-3+\mathbb{K}_{\bar{b}c}\right)=
\left(v'-3+\mathbb{K}_{\bar{b}c}\right)\left(u'-\mathbb{P}_{ac}\right)\left(u'-v'-1+\mathbb{K}_{a\bar{b}}\right)
\end{equation}
which is just the YBE obeyed by the fused $R$-matrices (\ref{fusedRsets}), 
\begin{equation}
R_{a\bar{b}}(u'-v')R_{ac}(u')R_{\bar{b}c}(v')=R_{\bar{b}c}(v')R_{ac}(u')R_{a\bar{b}}(u'-v'),
\end{equation}
and thus we complete the demonstration of the equivalence.

\subsection{Fused $K^-$-matrix solutions in $V_{\bar{\mathbf{4}}}$}
Now we consider the fused RE (\ref{fre}) for $n=1,n'=3$ with the parameters $c_{\ud{1}}=0,c_{\ud{2}}=-1,c_{\ud{3}}=-2$:
\begin{equation}\label{eq9}
\begin{split}
&R_{1,(\ud{1},\ud{2},\ud{3})}^{0,(0,-1,-2)}(u-v)K^{-}_1(u)R_{(\ud{1},\ud{2},\ud{3}),1}^{(0,-1,-2),0}(u+v)K_{(\ud{1},\ud{2},\ud{3})}^{-,(0,-1,-2)}(v)\\
=&K_{(\ud{1},\ud{2},\ud{3})}^{-,(0,-1,-2)}(v) R_{1,(\ud{3},\ud{2},\ud{1})}^{0,(2,1,0)}(u+v) K^{-}_1(u) R_{(\ud{3},\ud{2},\ud{1}),1}^{(2,1,0),0}(u-v).
\end{split}
\end{equation}
Due to the P-symmetric property:
\begin{equation}
 R_{1,(\ud{3},\ud{2},\ud{1})}^{0,(2,1,0)}(u+v)=R_{(\ud{1},\ud{2},\ud{3}),1}^{(0,-1,-2),0}(u+v),\quad 
 R_{(\ud{3},\ud{2},\ud{1}),1}^{(2,1,0),0}(u-v)=R_{1,(\ud{1},\ud{2},\ud{3})}^{0,(0,-1,-2)}(u-v),
\end{equation} 
and the relations (\ref{eq5}) and (\ref{eq6}), we know the invariant subspace for the above equation (\ref{eq9}) is: $V_\textbf{4}\otimes B_{(123)}^{(0,-1,-2)}V^{\otimes 3}_{\textbf{4}}=V_\textbf{4}\otimes V_{\bar{\textbf{4}}}$. When evaluated in $V_\textbf{4}\otimes V_{\bar{\textbf{4}}}$, the fused RE (\ref{eq9}) becomes:
\begin{equation}\label{eqn4}
R_{\textbf{4}\bar{\textbf{4}}}(u-v)K^-_{\textbf{4}}(u)R_{\bar{\textbf{4}}\textbf{4}}(u+v)K^-_{\bar{\textbf{4}}}(v)
=K^-_{\bar{\textbf{4}}}(v)R_{\bar{\textbf{4}}\textbf{4}}(u+v)K^-_{\textbf{4}}(u)R_{\textbf{4}\bar{\textbf{4}}}(u-v).
\end{equation}
Next we will find the concrete form of $K^-_{\bar{\mathbf{4}}}(u)$, which is derived by imposing the restriction:
\begin{equation}
K^-_{\bar{\mathbf{4}}}(u)=K^{-,(0,-1,-2)}_{(123)}(u)\bigg|_{V_{\bar{\mathbf{4}}}}.
\end{equation}
The calculation relies on the intertwining relation (see \ref{interKm}):
\begin{equation}
K_{(123)}^{-,(0,-1,-2)}(u)B_{(123)}^{(0,-1,-2)}=B_{(123)}^{(0,-1,-2)} K_{(321)}^{-,(-2,-1,0)}(u),
\end{equation}
and also the detailed expression of $K_{(321)}^{-,(-2,-1,0)}(u)$, which is obtained from the decomposition relation (\ref{dKm1}) as:
\begin{equation}
K_{(321)}^{-,(-2,-1,0)}(u)=K^-_1(u) R_{21}(2u-1)R_{31}(2u-2)K^-_2(u-1)R_{32}(2u-3)K^-_3(u-2),
\end{equation}
where  $K^-(u)$ is the solution of the fundamental RE. Then, by the action of the fused $K_{(123)}^{-,(0,-1,-2)}(u)$ matrix on the canonical basis (\ref{4barbasis}) of $V_{\mathbf{4}}$ , we will obtain the matrix form of $K^-_{\bar{\mathbf{4}}}(u)$.

The explicit form of $K^-_{\bar{\mathbf{4}}}(u)$ depends on the concrete selection of the solutions for $K^-_{\mathbf{4}}(u)$ which take the general form: $K^-_{\mathbf{4}}(u)=\xi \mathbb{I}+u \mathbb{E}$, where $\xi$ is a constant and  $\mathbb{E}$ must be chosen from the following two categories: (a) $\mathbb{E}$ is diagonal and $\mathbb{E}^2=\mathbb{I}$; (b) $\mathbb{E}$ is strictly triangular and $\mathbb{E}^2=0$ \cite{Arnaudon:2004sd}. In the subsequent calculations, we will choose three prototypical $K^-_{\mathbf{4}}(u)$ matrices from these two categories as shown below \cite{Bai:2024qtg}:
\begin{eqnarray}
{\mbox{type I:}}\quad 
  K^-_{\mathbf{4}}(u)=\left(
  \begin{array}{cccc}
  \xi+u&0&0&0\\
  0&\xi-u&0&0\\
  0&0&\xi-u&0\\
  0&0&0&\xi-u
  \end{array}\right),
\end{eqnarray}
\begin{eqnarray}\label{eq10}
{\mbox{type II:}}\quad 
  K^-_{\mathbf{4}}(u)=\left(
\begin{array}{cccc}
\xi& u&&\\
0&\xi&&\\
&&\xi&u\\
&&0&\xi
\end{array}
\right),\qquad 
{\mbox{type III:}}\quad 
  K^-_{\mathbf{4}}(u)=\left(
\begin{array}{cccc}
\xi& u&&\\
0&\xi&&\\
&&\xi&0\\
&&0&\xi
\end{array}
\right).
\end{eqnarray}
\subsubsection{$K^-_{\bar{\mathbf{4}}}(u)$ from Type I solution of $K^-_{\textbf{4}}(u)$}
In this case, the action of $K^{-,(0,-1,-2)}_{(123)}(u)$ on the basis $\{\bar{e}_i,i=1,2,3,4\}$ are:
\begin{equation}
\begin{split}
&K^{(0,-1,-2)}_{(123)}(u)\bar{e}_1=\frac{2u(\xi-u+2)(\xi-u+1)(\xi-u)}{2u-3}\bar{e}_1\\
&K^{(0,-1,-2)}_{(123)}(u)\bar{e}_i=\frac{2u(\xi-u+2)(\xi-u+1)(\xi+u-2)}{2u-3}\bar{e}_i,\quad i=2,3,4.
\end{split}
\end{equation}
Thus, by discarding the common prefactor $\frac{2u(\xi-u+2)(\xi-u+1)}{2u-3}$,  we obtain: 
\begin{equation}\label{eqn1}
K^-_{\mathbf{\bar{4}}}(u)=\left(
  \begin{array}{cccc}
    \xi-u &  & &  \\
     & \xi+u-2 & &  \\
     &  & \xi+u-2 &  \\
     &  & & \xi+u-2 \\
  \end{array}
  \right).
\end{equation}
For the particular case $\xi=1$, we find $K^-_{\mathbf{\bar{4}}}(u)$ proportional to a constant matrix:
\begin{equation}
K^-_{\mathbf{\bar{4}}}(u)=\left(
  \begin{array}{cccc}
    1-u &  & &  \\
     & u-1 & &  \\
     &  & u-1 &  \\
     &  & & u-1 \\
  \end{array}
  \right)
  \sim\left(
  \begin{array}{cccc}
    -1 &  & &  \\
     & 1 & &  \\
     &  & 1 &  \\
     &  & & 1 \\
  \end{array}
  \right),
  \end{equation}
  which coincides with the $K^-_{\bar{\mathbf{4}}}(u)$ solution used in the work \cite{Bai:2019soy}.
\subsubsection{$K^-_{\bar{\mathbf{4}}}(u)$ from Type II solution of $K^-_{\textbf{4}}(u)$}
When starting from the type II non-diagonal  $K^-_{\textbf{4}}(u)$ solution in (\ref{eq10}), the calculation becomes rather involved due to the mixing of the basis vectors $\bar{e}_i$ under the action of $K^{(0,-1,-2)}_{(123)}(u)$. However, the final result turns out to be quite simple:
\begin{equation}
\begin{split}
&K^{(0,-1,-2)}_{(123)}(u)\bar{e}_1=\frac{2u\xi^2}{2u-3}\left[\xi \bar{e}_1-(u-1)\bar{e}_2\right],\\
&K^{(0,-1,-2)}_{(123)}(u)\bar{e}_2=\frac{2u\xi^3}{2u-3}\bar{e}_2,\\
&K^{(0,-1,-2)}_{(123)}(u)\bar{e}_3=\frac{2u\xi^2}{2u-3}\left[\xi \bar{e}_3-(u-1)\bar{e}_4\right],\\
&K^{(0,-1,-2)}_{(123)}(u)\bar{e}_4=\frac{2u\xi^3}{2u-3}\bar{e}_4.
\end{split}
\end{equation}
After omitting the common scalar prefactor $\frac{2u\xi^2}{2u-3}$, the matrix representation of $K^-_{\bar{\textbf{4}}}(u)$ takes the form:
\begin{equation}\label{eqn2}
K^-_{\mathbf{\bar{4}}}(u)=\left(
  \begin{array}{cccc}
    \xi &0 &0 &0  \\
     -u+1 & \xi &0 &0  \\
     0&0  & \xi &0  \\
     0&0  & -u+1 & \xi \\
  \end{array}
  \right).
  \end{equation}
\subsubsection{$K^-_{\bar{\mathbf{4}}}(u)$ from Type III solution of $K^-_{\textbf{4}}(u)$}
In the case of Type III $K_{\textbf{4}}(u)$ in (\ref{eq10}), the operator $K^{(0,-1,-2)}_{(123)}(u)$ likewise mixes the basis vectors $\bar{e}_1$ and $\bar{e}_2$. The complete results are :
\begin{equation}
\begin{split}
&K^{(0,-1,-2)}_{(123)}(u)\bar{e}_1=\frac{2u\xi^2}{2u-3}\left[\xi \bar{e}_1-(u-1)\bar{e}_2\right],\\
&K^{(0,-1,-2)}_{(123)}(u)\bar{e}_i=\frac{2u\xi^3}{2u-3}\bar{e}_i,\quad i=2,3,4,
\end{split}
\end{equation}
from which we obtain the matrix form of $K^-_{\bar{\textbf{4}}}(u)$ as:
\begin{equation}\label{eqn3}
K^-_{\mathbf{\bar{4}}}(u)=\left(
  \begin{array}{cccc}
    \xi &0 &0 &0  \\
     -u+1 & \xi &0 &0  \\
     0&0  & \xi &0  \\
     0&0  & 0 & \xi \\
  \end{array}
  \right).
  \end{equation}
  
As a remark, we emphasize that all three $K^-_{\textbf{4}}(u)$ solutions in (\ref{eq9}) and (\ref{eq10}) and their corresponding  $K^-_{\bar{\textbf{4}}}(u)$ solutions in (\ref{eqn1}),(\ref{eqn2}) and (\ref{eqn3}) must satisfy the RE relation (\ref{eqn4}) (with the $R$-matrices given in (\ref{eq:fusedR}) and (\ref{eqn5})).  We have verified this using \textit{Mathematica} program.

As another application of the $\mathfrak{su}(4)$ fusion procedure, we show in the appendix \ref{app1} that the $R$-matrix of the well-known $\mathfrak{so}(6)$ spin chain in super Yang-Mills theory \cite{Minahan:2002ve} can be derived from the antisymmetric fusion of the fundamental $\mathfrak{su}(4)$ $R$-matrices.

\section{Conclusion}

In this paper, we present a general fusion procedure for open $\mathfrak{gl}(N)$ spin chains. We construct boundary reflection $K^-$- and $K^+$-matrices with inhomogeneity parameters, as well as their associated fused reflection equations and dual fused reflection equations. From the intertwining relations between the boundary reflection matrices and the fusion operators, we obtain the invariant subspaces for the $K^-$ and $K^+$-matrices. When specific inhomogeneity parameters are chosen, we derive the boundary reflection matrices restricted to finite-dimensional irreducible representations of $\mathfrak{gl}(N)$. Using the fused reflection matrices and fused $R$-matrices, we construct the corresponding fused transfer matrices and demonstrate the commutativity of these fused transfer matrices for different spectral parameters, thus serving as generating functions for the conserved charges. We also compute explicit expressions for the fused transfer matrices and find that: when the auxiliary space is the total tensor product space, the fused transfer matrix can be expressed as a product of fundamental transfer matrices. When the auxiliary space is projected to a one-dimensional invariant subspace by the fusion operator, the fused transfer matrix factorizes into a product of quantum determinants. As a specific application of the fusion procedure, we investigate the ABJM spin chain model. We find that, when the fusion operator is taken to be the projection operator onto the anti-fundamental representation space of $\mathfrak{su}(4)$, the resulting fused $R$-matrix reduces precisely to the $R$-matrix of the ABJM model, thus demonstrating that the ABJM spin chain is essentially an $\mathfrak{su}(4)$ fusion model. Using fusion techniques, we also obtain solutions for three classes of $K^-$-matrices defined on the anti-fundamental representation space. For another application of $\mathfrak{su}(4)$ fusion procedure,  we prove that the well-known $\mathfrak{so}(6)$ $R$-matrix originating from SYM spin chain can be obtained by the antisymmetric fusion of $\mathfrak{su}(4)$ $R$-matrices.

We would like to point out several directions for future research. First, starting from more general $K^-$-matrix solutions in $\mathfrak{su}(4)$ fundamental representation space (e.g., those related to our solutions by similarity transformations), we could obtain much broader classes of reflection matrix solutions in anti-fundamental representation spaces of $\mathfrak{su}(4)$ through fusion procedure. Furthermore, beyond the ABJM model context, solutions to the reflection equations could be considered in other irreducible representation spaces of $\mathfrak{su}(4)$. Second, based on the boundary reflection matrices derived by the fusion procedure, we can investigate the corresponding Hamiltonians with open boundaries and explore physically relevant new systems. We may also study the integrable boundary states constructed from these novel boundary reflection matrices. Third, a more challenging task is to investigate the eigenspectra of these novel open spin chain systems. For general non-diagonal boundary reflection matrices (such as those obtained in this work), the ODBA method may be required. Therefore, we need to establish the hierarchy relations among the fused transfer matrices defined in different representation subspaces, which may be obtained from the branching properties of the fusion operators.

\section*{Acknowledgments}
Nan Bai would like to thank Mao-Zhong Shao for collaboration at the early stage of this work, and Hong L\"{u}, Jun-Bao Wu, and Jia-ju Zhang for valuable discussions. This work was supported by the National Natural Science Foundation of China (Grant No. 12165002).

\begin{appendix}
\section{$\mathfrak{so}(6)$ integrable spin chain as a $\mathfrak{su}(4)$ fusion model}\label{app1}
Since the $\mathfrak{so}(6)$ spinor representation is isomorphic to the antisymmetric representation of $\mathfrak{su}(4)$, we consider the $\mathfrak{su}(4)$  fused $R$-matrix of the following type:
\begin{equation}
R^{(c_1,c_2),(c_{\ud{1}},c_{\ud{2}})}_{(12),(\ud{1}\ud{2})}(u)\in {\rm{End}}(V^{\otimes 2}\otimes V^{\otimes 2}).
\end{equation}
By choosing $c_1=c_{\ud{1}}=0,c_2=c_{\ud{2}}=-1$, we focus on the antisymmetric invariant subspaces $\Lambda V^{\otimes 2}\otimes \Lambda V^{\otimes 2}$:
\begin{equation}
\Lambda V^{\otimes 2}\otimes \Lambda V^{\otimes 2}=B_{12}^{(0,-1)}B_{\ud{1}\ud{2}}^{(0,-1)}V^{\otimes 2}\otimes V^{\otimes 2}.
\end{equation}
The basis for subspace $\Lambda V^{\otimes 2}$ is chosen as:
\begin{equation}
\begin{split}
E_1=e_1\otimes e_2-e_2\otimes e_1=B_{12}^{(0,-1)}e_1\otimes e_2 ,\quad E_2=e_1\otimes e_3-e_3\otimes e_1=B_{12}^{(0,-1)}e_1\otimes e_3,\\
E_3=e_1\otimes e_4-e_4\otimes e_1=B_{12}^{(0,-1)}e_1\otimes e_4 ,\quad E_4=e_2\otimes e_3-e_3\otimes e_2=B_{12}^{(0,-1)}e_2\otimes e_3,\\
E_5=e_2\otimes e_4-e_4\otimes e_2=B_{12}^{(0,-1)}e_2\otimes e_4 ,\quad E_6=e_3\otimes e_4-e_4\otimes e_3=B_{12}^{(0,-1)}e_3\otimes e_4,
\end{split}
\end{equation}
where $\{e_i\}$ are standard basis vectors in $V_\mathbf{4}$. The $R$-matrix we are looking for is:
\begin{equation}
R_{\mathbf{66}}(u)=R^{(0,-1),(0,-1)}_{(12),(\ud{1}\ud{2})}(u)|_{\Lambda V^{\otimes 2}\otimes \Lambda V^{\otimes 2}}.
\end{equation}
We denote the basis of the tensor space $\Lambda V^{\otimes 2}\otimes \Lambda V^{\otimes 2}$ by $E_{ij}=E_i\otimes E_j$. Then we find:
\begin{eqnarray}
R_{\mathbf{66}}(u)E_{16}=E_{16}+\frac{2}{u(u-1)}E_{61}-\frac{1}{u-1}E_{25}-\frac{1}{u-1}E_{52}+\frac{1}{u-1}E_{34}+\frac{1}{u-1}E_{43},\label{nneq1}\\
R_{\mathbf{66}}(u)E_{61}=E_{61}+\frac{2}{u(u-1)}E_{16}-\frac{1}{u-1}E_{25}-\frac{1}{u-1}E_{52}+\frac{1}{u-1}E_{34}+\frac{1}{u-1}E_{43},\\
R_{\mathbf{66}}(u)E_{25}=E_{25}+\frac{2}{u(u-1)}E_{52}-\frac{1}{u-1}E_{16}-\frac{1}{u-1}E_{61}-\frac{1}{u-1}E_{34}-\frac{1}{u-1}E_{43},\\
R_{\mathbf{66}}(u)E_{52}=E_{52}+\frac{2}{u(u-1)}E_{25}-\frac{1}{u-1}E_{16}-\frac{1}{u-1}E_{61}-\frac{1}{u-1}E_{34}-\frac{1}{u-1}E_{43},\\
R_{\mathbf{66}}(u)E_{34}=E_{34}+\frac{2}{u(u-1)}E_{43}+\frac{1}{u-1}E_{16}+\frac{1}{u-1}E_{61}-\frac{1}{u-1}E_{25}-\frac{1}{u-1}E_{52},\\
R_{\mathbf{66}}(u)E_{43}=E_{43}+\frac{2}{u(u-1)}E_{34}+\frac{1}{u-1}E_{16}+\frac{1}{u-1}E_{61}-\frac{1}{u-1}E_{25}-\frac{1}{u-1}E_{52},
\end{eqnarray}
and $R_{\mathbf{66}}(u) E_{ij}=\frac{u-2}{u-1}E_{ij}-\frac{u-2}{u(u-1)}E_{ji}$ for all other cases. Then, for example, we rewrite (\ref{nneq1}) as:
\begin{equation}
R_{\mathbf{66}}(u)E_{16}=\frac{u-2}{u-1}E_{16}-\frac{u-2}{u(u-1)}E_{61}+\frac{1}{u-1}\left[E_{16}+E_{61}+(-E_{25})+(-E_{52})+E_{34}+E_{43}\right].
\end{equation}
Thus, by redefining the indices of the basis, we can readily obtain the matrix form of $R_{\mathbf{66}}(u)$ as:
\begin{equation}
R_{\mathbf{66}}(u)=u(u-2)\mathbb{I}-(u-2)\mathbb{P}+u\mathbb{K},
\end{equation}
where an overall factor of $u(u-1)$ has been omitted. The $R$-matrix of integrable $\mathfrak{so}(n)$ spin chain, as given in \cite{Minahan:2002ve}, is:
\begin{equation}
R_{12}(u)=\frac{1}{n-2}\left[u(2u+2-n)\mathbb{I}-(2u+2-n)\mathbb{P}+2u\mathbb{K}\right],
\end{equation}
\end{appendix}
which coincides with ours when $n=6$.

\end{document}